\documentclass[a4paper]{article}

\usepackage{amsmath}
\usepackage{amssymb}
\usepackage[english]{babel}
\usepackage[small,bf]{caption}
\usepackage{cite}
\usepackage[T1]{fontenc}
\usepackage{graphicx}
\usepackage[hidelinks]{hyperref}
\usepackage{IEEEtrantools}
\usepackage[utf8]{inputenc}
\usepackage{multirow}
\usepackage{pdflscape}
\usepackage{xcolor}

\title{Therapeutic target discovery using Boolean network attractors: improvements of kali}
\author{Arnaud Poret, Carito Guziolowski}
\date{\today}

\begin{document}

\maketitle

\vfill

\begin{abstract}
In a previous article, an algorithm for identifying therapeutic targets in Boolean networks modeling pathological mechanisms was introduced. In the present article, the improvements made on this algorithm, named kali, are described. These improvements are i) the possibility to work on asynchronous Boolean networks, ii) a finer assessment of therapeutic targets and iii) the possibility to use multivalued logic. kali assumes that the attractors of a dynamical system, such as a Boolean network, are associated with the phenotypes of the modeled biological system. Given a logic-based model of pathological mechanisms, kali searches for therapeutic targets able to reduce the reachability of the attractors associated with pathological phenotypes, thus reducing their likeliness. kali is illustrated on an example network and used on a biological case study. The case study is a published logic-based model of bladder tumorigenesis from which kali returns consistent results. However, like any computational tool, kali can predict but can not replace human expertise: it is a supporting tool for coping with the complexity of biological systems in the field of drug discovery.
\end{abstract}

\vfill

\noindent Copyright 2016-2018 Arnaud Poret

\bigskip

\noindent This article is licensed under the Creative Commons Attribution-NonCommercial-NoDerivatives 4.0 International License. To view a copy of this license, visit \url{https://creativecommons.org/licenses/by-nc-nd/4.0/}.

\bigskip

\noindent \texttt{arnaud.poret@gmail.com} (corresponding author)\\
\texttt{carito.guziolowski@ls2n.fr}\\
LS2N, UMR 6004\\
Nantes, France

\newpage

\tableofcontents

\newpage

\section{Introduction}

In a previous article, an algorithm for \textit{in silico} therapeutic target discovery was presented in its first version \cite{poret2014silico}. In the present article, the improvements made on this algorithm, named kali, are described. The complete background was introduced in the previous article whose some important concepts are recalled in \hyperref[brief]{\texttt{Appendix 1}} page \pageref{brief}.

kali still belongs to the logic-based modeling formalism \cite{le2015quantitative,wynn2012logic,morris2010logic}, mainly Boolean networks \cite{albert2014boolean,wang2012boolean}, and keeps its original goal: searching for therapeutic interventions aimed at healing a supplied pathologically disturbed biological network. Such a network is intended to model the biological mechanisms of a studied disease and is on what kali operates. Therapeutic interventions are combinations of targets, these combinations being named bullets. Targets are network components, such as enzymes or transcription factors, and can be subjected to inhibition or activation. This is what bullets specify: which targets and which actions to apply on them.

The pivotal assumption on which kali is based postulates that the attractors of a dynamical system, such as a Boolean network, are associated with the phenotypes of the modeled biological system. In other words: attractors model phenotypes \cite{jaeger2014bioattractors}. This assumption was successfully applied in several works \cite{cho2016attractor,gan2016analysis,davila2015modeling,crespo2013detecting,fumia2013boolean,cheng2013biomolecular,creixell2012navigating} and makes sense since the steady states of a dynamical system, the attractors, should mirror the steady states of the modeled biological system, the phenotypes.

In the mean time, various works using logical modeling with application in therapeutic innovation were published. An example is the work of Hyunho Chu and colleagues \cite{chu2015precritical}. They built a molecular interaction network involved in colorectal tumorigenesis and studied its dynamics, particularly its attractors and their basins, with stochastic Boolean modeling. They highlighted what they termed the flickering, that is the displacement of the system from one basin to another one due to stochastic noise. They suggested that the flickering is involved in pushing the system from a physiological state to a pathological one during colorectal tumorigenesis.

Concerning kali, three improvements were done: i) adding the possibility to work with asynchronous Boolean networks, ii) implementing a finer assessment of therapeutic targets and iii) adding the possibility to use multivalued logic. The technical features resulting from these improvements are illustrated on a simple example network while their biological significance is assessed on a case study, namely a published logic-based model of bladder tumorigenesis \cite{remy2015modeling}.

\subsection{Handling asynchronous updating}

\label{async} To compute the behavior of a discrete dynamical system, such as a Boolean network, its variables have to be iteratively updated. These iterative updates can be made synchronously or not \cite{garg2008synchronous}. If all the variables are simultaneously updated at each iteration then the network is synchronous, otherwise it is asynchronous. Compared to an asynchronous updating, the synchronous one is easier to compute. However, when the dynamics of a biological network is computed synchronously, it is assumed that all its components evolve simultaneously, an assumption which can be inappropriate according to what is modeled.

The asynchronous updating is frequently built so that one randomly selected variable is updated at each iteration. This allows to capture two important features: i) biological entities do not necessarily evolve simultaneously and ii) noise due to randomness can affect when biological interactions take place \cite{szekely2014stochastic,buiatti2013randomness,ullah2010stochastic}. This is particularly true at the molecular scale, such as with signaling pathways, where macromolecular crowding and Brownian motion can impact the firing of biochemical reactions \cite{rivas2016macromolecular}.

Therefore, the choice between a synchronous and an asynchronous updating may depend on the model, the computational resources and the acceptability of synchrony. Knowing that the luxury is to have the choice, kali can now use synchronous and asynchronous updating.

\subsection{Managing basin sizes for therapeutic purpose}

Until now, kali requires therapeutic bullets to remove all the attractors associated with pathological phenotypes, here named pathological attractors. This criterion for selecting therapeutic bullets is somewhat drastic. A smoother criterion should enable to consider more targeting strategies and then more possibilities for counteracting diseases. However, it could also unravel less effective therapeutic bullets, but being too demanding potentially leads to no results and the loss of nonetheless interesting findings.

The therapeutic potential of bullets could be assessed by estimating their ability at reducing the size of the pathological basins, namely the basins of pathological attractors. This criterion is more permissive since therapeutic bullets no longer have to necessarily remove the pathological attractors. Reducing the size of a pathological basin renders the corresponding pathological attractor less reachable and then the associated pathological phenotype less likely. This new criterion includes the previous one: removing an attractor means reducing its basin to the empty set. Consequently, therapeutic bullets obtainable with the previous criterion are still obtainable.

\subsection{Extending to multivalued logic}

One of the main limitations of Boolean models is that their variables can take only two values, which can be too simplistic in some cases. Depending on what is modeled, such as activity level of enzymes or abundance of gene products, considering more than two levels can be better. Without leaving the logic-based modeling formalism, one solution is to extend Boolean logic to multivalued logic \cite{rescher1968many}.

With multivalued logic, a finite number $h$ of values in the interval of real numbers $[0;1]$ is used, thus allowing variables to model more than two levels. For example, the level $0.5$ can be introduced to model partial activation of enzymes or moderate concentration of gene products.

\section{Methods}

\subsection{Additional definitions}

In addition to the background introduced in the previous article \cite{poret2014silico} and briefly recalled in \hyperref[brief]{\texttt{Appendix 1}} page \pageref{brief}, here are some supplementary definitions:
\begin{itemize}
\item \textbf{physiological state space}: the state space $S_{physio}$ of the physiological variant
\item \textbf{pathological state space}: the state space $S_{patho}$ of the pathological variant
\item \textbf{testing state space}: the state space $S_{test}$ of the pathological variant under the effect of a bullet
\item \textbf{physiological basin}: the basin $B_{physio,i}$ of a physiological attractor $a_{physio,i}$
\item \textbf{pathological basin}: the basin $B_{patho,i}$ of a pathological attractor $a_{patho,i}$
\item \textbf{$n$-bullet}: a bullet made of $n$ targets
\end{itemize}

\subsection{Handling asynchronous updating}

To incorporate asynchronous updating, the corresponding algorithms coming from BoolNet were implemented into kali. BoolNet is an R \cite{R} package for generation, reconstruction and analysis of Boolean networks \cite{mussel2010boolnet}. Asynchronous updating is implemented so that one randomly selected variable is updated at each iteration. This random selection is made according to a uniform distribution and implies that the network is no longer deterministic. To do so, given a Boolean network, BoolNet uses the three following functions:
\begin{itemize}
\item \textbf{AsynchronousAttractorSearch}: this function computes the attractor set of a supplied Boolean network by using the two following functions
\item \textbf{ForwardSet}: this function computes the forward reachable set (see below) of a state and considers it as a candidate attractor
\item \textbf{ValidateAttractor}: this function checks if a forward reachable set is a terminal strongly connected component (terminal SCC, see below), that is an attractor
\end{itemize}

The forward reachable set $Fwd_{\boldsymbol{x}} \subset S$ of a state $\boldsymbol{x} \in S$ is the set made of the states reachable from $\boldsymbol{x}$, including $\boldsymbol{x}$ itself. A terminal SCC is a set $tSCC \subset S$ made of the forward reachable sets of its states: $\forall \boldsymbol{x} \in tSCC$, $Fwd_{\boldsymbol{x}} \subset tSCC$. As a consequence, when a terminal SCC is reached, the system can not escape it: this is an attractor in the sense of asynchronous Boolean networks \cite{saadatpour2010attractor}.

\label{kmax} Asynchronous Boolean networks with random updating are not deterministic: their attractors are no longer deterministic sequences of states, namely cycles, but terminal SCCs. To find such an attractor, a long random walk is performed in order to reach an attractor with high probability. This candidate attractor is then validated, or not, by checking if it is a terminal SCC.

\subsection{Managing basin sizes for therapeutic purpose}

To implement the new criterion for selecting therapeutic bullets, kali considers a bullet as therapeutic if it increases the union of the physiological basins $\bigcup B_{physio,i}$ in the testing state space $S_{test}$ without creating \textit{de novo} attractors. Knowing that for kali an attractor is either physiological or pathological, increasing $\bigcup B_{physio,i}$ is equivalent to decreasing $\bigcup B_{patho,i}$.

The goal is to increase the physiological part of the pathological state space, or equivalently to decrease its pathological part. Consequently, a pathologically disturbed biological network receiving such a therapeutic bullet tends to, but not necessarily reaches, an overall physiological behavior.

However, as with the previous criterion, it does not ensure that all the physiological attractors are preserved. \textit{A fortiori}, it does not ensure that their basin remains unchanged. It means that a therapeutic bullet can also alter the reachability of the physiological attractors. Nevertheless, as with the previous criterion, this is a matter of choice between a therapeutic bullet or no bullet at all.

The therapeutic potential of a bullet is expressed by its gain. It is displayed as follows:
\begin{small}
\begin{IEEEeqnarray*}{c}
x\% \to y\% \text{ with } x=100 \cdot \frac{|\bigcup B_{physio,i}|}{|S_{patho}|} \text{ and } y=100 \cdot \frac{|\bigcup B_{physio,i}|}{|S_{test}|}
\end{IEEEeqnarray*}
\end{small}
expressed in percents. Therefore, in order to increase the physiological part of the pathological state space, a therapeutic bullet has to make $y \geq x$.

Note that $y=x$ is allowed. In this particular case, it is conceivable that the size of several pathological basins changed while the size of their union did not. In other words, the composition of the pathological part changed while its size did not. It can be therapeutic if, for example, the basin of a weakly pathological attractor increases at the expense of the basin of a heavily pathological attractor.

\label{threshold} The increase of the physiological part of the pathological state space can be subjected to a threshold $\delta$: $y \geq x$ becomes $y-x \geq \delta$. As $x$ and $y$, $\delta$ is expressed in percents of the state space. This threshold is introduced to allow the stringency of kali to be tuned. By the way, using this threshold also decreases the probability to obtain misassessed therapeutic bullets due to roundoff errors, or sampling errors when the state space is too big to compute trajectories from each of the possible states.

A therapeutic bullet as defined by the previous criterion, namely which removes all the pathological attractors, makes \textit{de facto} $\bigcup B_{physio,i}=100\%$ of $S_{test}$. As already mentioned, the previous criterion is included in this new one: therapeutic bullets obtainable with the former are also obtainable with the latter.

It must be pointed out that the current implementation of the method described in this article, namely kali, computes basin sizes by counting the number of initial states leading to a given attractor. If these initial states are a subset of the state space then basin sizes are estimations. Moreover, if an asynchronous updating is used then the system is not deterministic, implying that an initial state can lead to more than one attractor. Consequently, in those cases, basin sizes and therapeutic gains are estimations also subjected to random variations.

In other words, concerning the calculation of basin sizes, the current implementation of kali is more an attractor reachability estimation than a true basin size calculation. Nevertheless, speaking in terms of basins is kept in order to better comply with the underlaying method, independently of its implementation which is subjected to further improvements.

\subsection{Extending to multivalued logic}

Extending to multivalued logic requires suitable operators to be introduced. One solution is to use an implementation of the Boolean operators which also works with multivalued logic, just as the Zadeh operators. These operators are a generalization of the Boolean ones proposed for fuzzy logic by its pioneer Lotfi Zadeh \cite{zadeh1965fuzzy}. Their formulation is:
\begin{small}
\begin{IEEEeqnarray*}{r C l}
x \land y&=&min(x,y)\\
x \lor y&=&max(x,y)\\
\lnot x&=&1-x
\end{IEEEeqnarray*}
\end{small}

With a $h$-valued logic, the size of the $n$-dimensional state space is $h^{n}$, bringing more computational difficulties than with Boolean logic. The same applies to the testable bullets since there are $h^{r}$ possible modality arrangements and then $(n! \cdot h^{r})/(r! \cdot (n-r)!)$ possible bullets, where $r$ is the number of targets per bullet (see below).

As introduced in the previous article \cite{poret2014silico} and recalled in \hyperref[brief]{\texttt{Appendix 1}} page \pageref{brief}, a bullet is a couple $(c_{targ},c_{moda})$ where $c_{targ}=(targ_{1},\ldots,targ_{r})$ is a combination without repetition of $r$ nodes and $c_{moda}=(moda_{1},\ldots,moda_{r})$ is an arrangement with repetition of $r$ perturbations, here termed modalities. $moda_{i}$ is intended to be applied on $targ_{i}$.

To illustrate how kali works with multivalued logic without overloading it, a $3$-valued logic is used with $\lbrace 0,0.5,1 \rbrace$ as domain of value: $x_{i} \in \lbrace 0,0.5,1 \rbrace$. $0$ and $1$ have the same meaning as with Boolean logic. $0.5$ is an intermediate truth degree which can be interpreted as an intermediate level of activity\slash abundance depending on what the variables refer to. By the way, $S=\lbrace 0,0.5,1 \rbrace^{n}$ and $moda_{i} \in \lbrace 0,0.5,1 \rbrace$.

\subsection{Example network}

\begin{figure*}[!hbp]
\begin{center}
\includegraphics[width=0.5\textwidth]{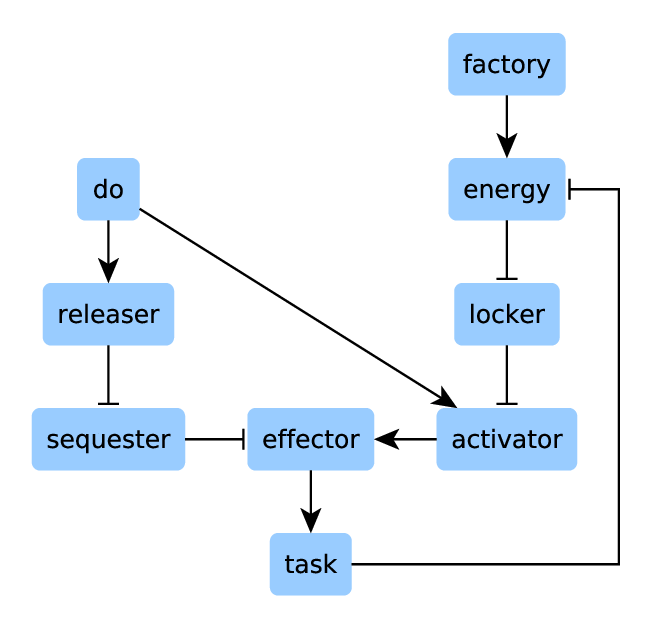}
\caption[The example network]{\label{network} This network, running in a fictive cell, controls the execution of a task according to two inputs: i) the do instruction, which tells the task to be performed, and ii) energy supply. The task consumes energy and must be prevented if no energy is available, even if the do instruction is sent. The task is initiated by an effector, which is maintained inactive by a sequester. The do instruction activates a releaser which suppresses the sequestering activity of the sequester, thus releasing the effector. However, to initiate the task and in addition to be released, the effector has also to be activated by an activator. When released and activated, the effector initiates the task. To ensure that the task is performed only if energy is available, a locker maintains the activator in an inactive state if there is no energy, even if the do instruction is sent. Concerning the factory, it supplies energy.}
\end{center}
\end{figure*}

To conveniently illustrate the technical features resulting from the improvements made on kali, a simple and fictive example network is used. A biological case study is then proposed to address a concrete case, namely a published logic-based model of bladder tumorigenesis \cite{remy2015modeling}. The example network is depicted in \hyperref[network]{\texttt{Figure \ref*{network}}} page \pageref{network}.

Among the three improvements made on kali, only the asynchronous updating and the management of basin sizes are illustrated. Multivalued logic is a straightforward extension of the Boolean case and is illustrated in \hyperref[multivalued]{\texttt{Appendix 2}} page \pageref{multivalued}. Below are the Boolean equations encoding the example network, also available in text format in the supporting file \texttt{example\_equations.txt}:
\begin{small}
\begin{IEEEeqnarray*}{r C l}
do&=&do\\
factory&=&factory\\
energy&=&factory \lor (energy \land \lnot task)\\
locker&=&\lnot energy\\
releaser&=&do\\
sequester&=&\lnot releaser\\
activator&=&do \land \lnot locker\\
effector&=&activator \land \lnot sequester\\
task&=&effector
\end{IEEEeqnarray*}
\end{small}

The do instruction and the factory are the two inputs: they are constant and therefore equal to themselves. The equation of energy tells that energy is present if the factory is active, even when the task is running: the factory has a sufficient production capacity. However, if the factory is not active then energy disappears as soon as the task is initiated. Concerning the activator and the effector, their equations tell that their respective inhibitor takes precedence: whatever is the state of the other nodes, if the inhibitor is active then the target is not.

The physiological variant $\boldsymbol{f}_{physio}$ is the network as is. The pathological variant $\boldsymbol{f}_{patho}$ is the network plus a constitutive inactivation of the locker: the execution of the task no longer considers if energy is available. Consequently, $f_{locker}$ becomes $locker=0$ in $\boldsymbol{f}_{patho}$.

\subsection{Case study: bladder tumorigenesis}

The case study consists in running kali on a logic-based model of bladder tumorigenesis published by Elisabeth Remy and colleagues \cite{remy2015modeling}. Elisabeth Remy and colleagues have built an influence network linking three extracellular input signals and one intracellular input event to three cellular output phenotypes.

The three extracellular input signals are growth stimulations, represented by the $EGFRstimulus$ and $FGFR3stimulus$ parameters, and growth inhibitions, mainly modeling TGF-$\beta$ effects and represented by the $GrowthInhibitors$ parameter. The intracellular input event is DNA damage, represented by the $DNAdamage$ parameter. The three cellular output phenotypes are proliferation, growth arrest and apoptosis. The model integrates downstream effectors of growth factor receptors such as Ras and PI3K, growth inhibitors such as p14ARF and p16INK4a, and regulators of the cell cycle such as cyclinD1, E2F3 and pRb.

Some variables are ternary: they can take three possible values in order to account for different effects depending on the activation level. These three possible values are $0$ and $1$ as in the Boolean case, plus the additional level $2$. As in the model implementation performed by Elisabeth Remy and colleagues, these ternary variables are translated into pairs of Boolean variables: one Boolean variable per activation level, namely level $1$ and level $2$.

For example and according to the model, in its normal expression level (level $1$, $E2F1=1$) the transcription factor E2F1 stimulates the expression of genes supporting the cell cycle. However, when over-expressed (level $2$, $E2F1=2$) E2F1 stimulates the expression of genes supporting apoptosis. Consequently, this ternary variable is translated into the pair of Boolean variables $E2F1_{lvl1}$ and $E2F1_{lvl2}$:
\begin{small}
\begin{IEEEeqnarray*}{r C l}
E2F1=1&\Leftrightarrow&E2F1_{lvl1}=1\\
E2F1=2&\Leftrightarrow&E2F1_{lvl2}=1
\end{IEEEeqnarray*}
\end{small}

The variable modeling the output phenotype $Apoptosis$ is one of these ternary variables. The goal of Elisabeth Remy and colleagues was to relate apoptosis to its trigger: p53-dependent apoptosis ($Apoptosis_{lvl1}$) and E2F1-dependent apoptosis ($Apoptosis_{lvl2}$). However, in the present case study, only the cell fate matters. These two trigger-dependent apoptosis are therefore merged into one equation:
\begin{small}
\begin{IEEEeqnarray*}{r C l}
Apoptosis&=&Apoptosis_{lvl1} \lor Apoptosis_{lvl2}
\end{IEEEeqnarray*}
\end{small}

\label{phenotypes} Since the four inputs of the model are parameters, their respective value are directly injected into the concerned equations so that no equations are dedicated to them, thus reducing computational requirements. Again to reduce computational requirements and knowing that the three output phenotypes are readouts not influencing other variables, their corresponding equation are put out of the model and evaluated from the returned attractors once the run terminated:
\begin{small}
\begin{IEEEeqnarray*}{r C l}
Proliferation&=&CyclinE1 \lor CyclinA\\
GrowthArrest&=&p21CIP \lor RB1 \lor RBL2\\
Apoptosis&=&TP53 \lor E2F1_{lvl2}
\end{IEEEeqnarray*}
\end{small}

\begin{figure*}[!hbp]
\begin{center}
\includegraphics[width=1\textwidth]{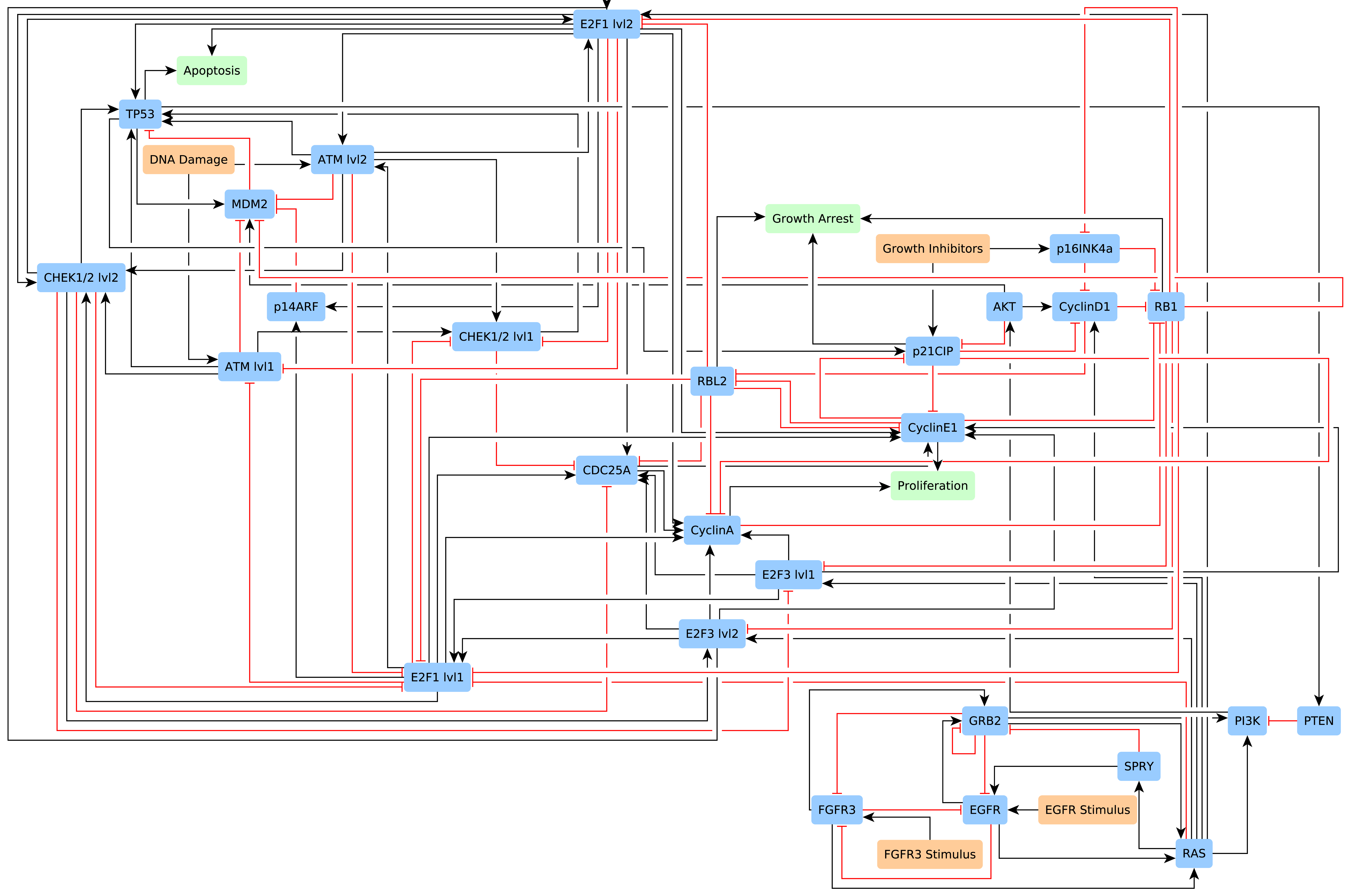}
\caption[The case study]{\label{bladder} A network-based representation of the case study used to assess kali on a concrete case. As explained in the text, it is derived from a published logic-based model of bladder tumorigenesis \cite{remy2015modeling}. Nodes represent Boolean variables while edges indicate positive (black) and negative (red) influences. The input signals\slash events growth stimulations, growth inhibitions and DNA damage are in red while the output phenotypes proliferation, growth arrest and apoptosis are in green.}
\end{center}
\end{figure*}

Altogether, the above described adaptations made on the model of bladder tumorigenesis by Elisabeth Remy and colleagues give a case study of $27$ Boolean equations. These equations are listed in \hyperref[equations]{\texttt{Appendix 4}} page \pageref{equations}, also available in text format in the supporting file \texttt{bladder\_equations.txt}. A network-based representation is shown in \hyperref[bladder]{\texttt{Figure \ref*{bladder}}} page \pageref{bladder}.

The physiological variant $\boldsymbol{f}_{physio}$ is the model as is. The pathological variant $\boldsymbol{f}_{patho}$ is the model plus a deletion of the tumor suppressor gene CDKN2A, as observed in bladder cancers \cite{cairns1994homozygous,meeks2016genomic}. Note that the CDKN2A gene encodes two growth inhibitors: p14ARF and p16INK4a. Consequently, the equations modeling these two variables become $p14ARF=0$ and $p16INK4a=0$ in $\boldsymbol{f}_{patho}$.

\subsection{Implementation, code availability, license}

kali is implemented in Go \cite{Go}, tested with Go version go1.9.2 linux\slash amd64 under Arch Linux \cite{ArchLinux}. kali is licensed under the GNU General Public License \cite{GPL} and freely available on GitHub at \url{https://github.com/arnaudporet/kali}. The core of kali in pseudocode can be found in \hyperref[pseudocode]{\texttt{Appendix 3}} page \pageref{pseudocode}.

\section{Results}

\subsection{Example network}

\subsubsection{Attractor sets}

The example network is computed asynchronously over the whole state space, namely $512$ possible initial states, using Boolean logic. As explained in the \hyperref[kmax]{\texttt{Methods}} section page \pageref{kmax}, the asynchronous attractor search uses long random walks to reach candidate attractors with high probability, and then checks if they are indeed true attractors. Owing to the small size of the example network, the length $max_{k}$ of these random walks is set to $1\ 000$ steps. With larger state spaces, random walks should be longer to reach candidate attractors with high probability.

The resulting attractors can be studied along four variables: the do instruction, the factory, the locker and the task. It is possible for energy to be present without a running factory in the initial conditions. In this case, if the do instruction is sent then energy is consumed by the task but not remade by the factory. With the physiological variant, the locker is expected to stop the task. However, with the pathological variant where the locker is disabled, an abnormal behavior is expected. Below are the computed attractors:
\begin{itemize}
\item $A_{physio}$:
\begin{small}
\begin{center}
\begin{tabular}{c|c|c|c|c|c|c}
attractor&basin ($\%$ of $S_{physio}$)&$do$&$factory$&$energy$&$locker$&$task$\\ \hline
$a_{physio1}$&$17.8\%$&$0$&$0$&$0$&$1$&$0$\\
$a_{physio2}$&$7.2\%$&$0$&$0$&$1$&$0$&$0$\\
$a_{physio3}$&$25\%$&$0$&$1$&$1$&$0$&$0$\\
$a_{physio4}$&$25\%$&$1$&$0$&$0$&$1$&$0$\\
$a_{physio5}$&$25\%$&$1$&$1$&$1$&$0$&$1$
\end{tabular}
\end{center}
\end{small}
\item $A_{patho}$:
\begin{small}
\begin{center}
\begin{tabular}{c|c|c|c|c|c|c}
attractor&basin ($\%$ of $S_{patho}$)&$do$&$factory$&$energy$&$locker$&$task$\\ \hline
$a_{patho1}$&$18.4\%$&$0$&$0$&$0$&$0$&$0$\\
$a_{physio2}$&$6.6\%$&$0$&$0$&$1$&$0$&$0$\\
$a_{physio3}$&$25\%$&$0$&$1$&$1$&$0$&$0$\\
$a_{patho2}$&$25\%$&$1$&$0$&$0$&$0$&$1$\\
$a_{physio5}$&$25\%$&$1$&$1$&$1$&$0$&$1$
\end{tabular}
\end{center}
\end{small}
\end{itemize}

With the physiological variant, the behavior is as expected: the task runs only if the do instruction is sent and only if the factory can remade the consumed energy. With the pathological variant, two pathological phenotypes represented by $a_{patho1}$ and $a_{patho2}$ appear. $a_{patho1}$ is pathological because the locker is inactive while there is no available energy. However, it is weakly pathological since the do instruction is not sent: there is no task to stop, an operational locker is not mandatory.

In contrast, $a_{patho2}$ is heavily pathological because an operational locker is required to stop the task in absence of energy supply. In the fictive cell bearing this example network, $a_{patho2}$ could drain all its energy content, thus bringing it to thermodynamical death. Moreover, $a_{patho2}$ should not be neglected since its basin occupies $25\%$ of the pathological state space.

\subsubsection{Therapeutic bullets}

Bullets are assessed for their therapeutic potential on the pathological variant $\boldsymbol{f}_{patho}$ according to the new criterion: decreasing the size of the pathological basins $B_{patho,i}$. All the bullets made of one to two targets are tested with a threshold of $5\%$.

Choosing a threshold can appear somewhat arbitrary. It tells that if the physiological part $\bigcup B_{physio,i}$ in the pathological state space $S_{patho}$ occupies $x\%$ of it, then to be therapeutic a bullet has to bring this value above $(x+5)\%$ in the testing state space $S_{test}$. Therefore, the increases below this threshold are considered not significant by kali. Even the choice of using a threshold can be arbitrary, as discussed in the \hyperref[threshold]{\texttt{Methods}} section page \pageref{threshold}.

Knowing that $\bigcup B_{physio,i}=56.6\%$ of $S_{patho}$, with a threshold of $5\%$ the $1,2$-bullets have to make $\bigcup B_{physio,i} \geq (56.6+5)\%=61.6\%$ of $S_{test}$ to be considered therapeutic. Below are the returned therapeutic bullets:

\begin{landscape}

\begin{itemize}
\item $1$-therapeutic bullets:
\begin{small}
\begin{center}
\begin{tabular}{r|r c r|r r r r r r r}
\multicolumn{1}{c}{bullet}&\multicolumn{3}{|c|}{gain}&$B_{physio1}$&$B_{physio2}$&$B_{physio3}$&$B_{physio4}$&$B_{physio5}$&$B_{patho1}$&$B_{patho2}$\\ \hline
$do[0]$&$56.6\%$&$\to$&$64.4\%$&$0\%$&$14.4\%$&$50\%$&$0\%$&$0\%$&$35.5\%$&$0\%$\\
$factory[1]$&$56.6\%$&$\to$&$100\%$&$0\%$&$0\%$&$50\%$&$0\%$&$50\%$&$0\%$&$0\%$
\end{tabular}
\end{center}
\end{small}
\item $2$-therapeutic bullets:
\begin{small}
\begin{center}
\begin{tabular}{r r|r c r|r r r r r r r}
\multicolumn{2}{c}{bullet}&\multicolumn{3}{|c|}{gain}&$B_{physio1}$&$B_{physio2}$&$B_{physio3}$&$B_{physio4}$&$B_{physio5}$&$B_{patho1}$&$B_{patho2}$\\ \hline
$do[0]$&$factory[1]$&$56.6\%$&$\to$&$100\%$&$0\%$&$0\%$&$100\%$&$0\%$&$0\%$&$0\%$&$0\%$\\
$do[1]$&$factory[1]$&$56.6\%$&$\to$&$100\%$&$0\%$&$0\%$&$0\%$&$0\%$&$100\%$&$0\%$&$0\%$\\
$do[0]$&$energy[1]$&$56.6\%$&$\to$&$100\%$&$0\%$&$50\%$&$50\%$&$0\%$&$0\%$&$0\%$&$0\%$\\
$do[0]$&$locker[0]$&$56.6\%$&$\to$&$64.1\%$&$0\%$&$14.1\%$&$50\%$&$0\%$&$0\%$&$35.9\%$&$0\%$\\
$do[0]$&$releaser[0]$&$56.6\%$&$\to$&$62.9\%$&$0\%$&$12.9\%$&$50\%$&$0\%$&$0\%$&$37.1\%$&$0\%$\\
$do[0]$&$sequester[1]$&$56.6\%$&$\to$&$62.5\%$&$0\%$&$12.5\%$&$50\%$&$0\%$&$0\%$&$37.5\%$&$0\%$\\
$do[0]$&$activator[0]$&$56.6\%$&$\to$&$64.8\%$&$0\%$&$14.8\%$&$50\%$&$0\%$&$0\%$&$35.2\%$&$0\%$\\
$do[0]$&$effector[0]$&$56.6\%$&$\to$&$67.8\%$&$0\%$&$17.8\%$&$50\%$&$0\%$&$0\%$&$32.2\%$&$0\%$\\
$do[0]$&$task[0]$&$56.6\%$&$\to$&$73.2\%$&$0\%$&$23.2\%$&$50\%$&$0\%$&$0\%$&$26.8\%$&$0\%$\\
$factory[1]$&$energy[1]$&$56.6\%$&$\to$&$100\%$&$0\%$&$0\%$&$50\%$&$0\%$&$50\%$&$0\%$&$0\%$\\
$factory[1]$&$locker[0]$&$56.6\%$&$\to$&$100\%$&$0\%$&$0\%$&$50\%$&$0\%$&$50\%$&$0\%$&$0\%$
\end{tabular}
\end{center}
\end{small}
\end{itemize}
where $x[y]$ means that the variable $x$ has to be set to the value $y$. For example, the therapeutic bullet $do[0]\ factory[1]$ suggests to abolish the do instruction while maintaining the factory active.

\end{landscape}

All the returned therapeutic bullets not removing all the pathological attractors exhibit the ability to suppress the basin of $a_{patho2}$ while increasing the one of $a_{patho1}$. Certainly, removing all the pathological attractors should be better, but knowing the $a_{patho2}$ is more pathological than $a_{patho1}$, such therapeutic bullets can nevertheless be interesting. With the previous criterion, namely removing all the pathological attractors, these therapeutic bullets are not obtainable, thus highlighting fewer therapeutic strategies.

Some of the found therapeutic bullets enable physiological attractors required by the pathological variant to react properly to the do instruction. For example, the therapeutic bullet $factory[1]$ enables $a_{physio3}$ and $a_{physio5}$, corresponding respectively to ``no do, no task'' and ``do the task, energy supply''. However, the remaining of the therapeutic bullets, such as $do[0]\ releaser[0]$ or $do[1]\ factory[1]$, either disable or force the do instruction, thus either suppressing or forcing the task. A network unable at performing the task or, at the opposite, permanently doing it may not be therapeutically interesting, even if energy is supplied.

None of the found therapeutic bullets suggest to reverse the constitutive inactivation of the locker. This highlights that applying the opposite action of a pathological disturbance is not necessarily a therapeutic solution, which can appear counterintuitive. This is because biological entities subjected to pathological disturbances belong to complex networks exhibiting behaviors which can not be predicted by mind \cite{koutsogiannouli2013complexity,koch2012modular}. In such context, computational tools and their growing computing capabilities can help owing to their integrative power \cite{jessica2016multi,walpole2013multiscale,fisher2010executable,voit2008steps,fischer2008mathematical}.

Also, none of the found therapeutic bullets allow the recovery of all the physiological attractors: there are no golden bullets. In a general manner, the components of biological networks should be able to take several states, such as enzymes which should be active when suitable. Consequently, healing a pathologically disturbed biological network by maintaining some of its components in a particular state should not allow the recovery of a complete and healthy behavior. This is a limitation of the method implemented in kali.

This limitation is common in biomedicine while not necessarily being an issue. For example, statins are well known lipid-lowering drugs widely used in cardiovascular diseases with proven benefits \cite{mihos2014cardiovascular,schooling2014statins}. They inhibit an enzyme, the HMG-CoA reductase, and they do it constantly, just as the targets are modulated in the therapeutic bullets returned by kali. The HMG-CoA reductase is component of a complex metabolic network and maintaining it in an inhibited state should not allow this network to run properly, maybe causing some adverse effects. Nevertheless, such as with all drugs, this is a matter of benefit-risk ratio.

All of this is to point that there are no perfect strategies for counteracting diseases and that computational tools, such as kali, can help scientists but can certainly not replace their expertise. Human expertise is mandatory to assess the returned predictions according to a concrete setting, and ultimately to take decisions.

\subsection{Case study: bladder tumorigenesis}

\subsubsection{Attractor sets}

The case study is computed asynchronously using Boolean logic. The state space being quite big with $134\ 217\ 728$ possible states, to compute an attractor set kali performs random walks starting from $1\ 000$ randomly selected initial states. A bigger state space also requires these random walks to be longer in order to reach candidate attractors with high probability. The length $max_{k}$ of the random walks is then increased to $10\ 000$ steps.

The $4$ input parameters of the model are tuned to simulate a biological situation where undamaged cells receive both growth stimulating and growth inhibiting signals from their environment:
\begin{small}
\begin{IEEEeqnarray*}{r C l}
EGFRstimulus&=&1\\
FGFR3stimulus&=&1\\
GrowthInhibitors&=&1\\
DNAdamage&=&0
\end{IEEEeqnarray*}
\end{small}

This input configuration aims at predicting the possible responses of the model to opposite growth instructions. In a cancerous setting, it is desirable that the growth inhibiting signal takes precedence over the stimulating one. With the pathological variant where the two growth inhibitors p14ARF and p16INK4a are absent, this desired precedence might be compromised in favor of tumorigenesis, thus correlating with the observed CDKN2A gene deletion in bladder cancers \cite{cairns1994homozygous,meeks2016genomic}.

The phenotypes associated with the returned attractors are evaluated using their respective equation once the run terminated, as explained in the \hyperref[phenotypes]{\texttt{Methods}} section page \pageref{phenotypes}. Below are the computed attractors together with their phenotypes and basins, expressed in percents of the corresponding state space:
\begin{small}
\begin{center}
\begin{tabular}{r|c|c|c|c|c}
&\multicolumn{3}{|c}{$A_{physio}$}&\multicolumn{2}{|c}{$A_{patho}$}\\ \hline
name&$a_{physio1}$&$a_{physio2}$&$a_{physio3}$&$a_{physio1}$&$a_{patho1}$\\
basin&$10.7\%$&$74.5\%$&$14.8\%$&$65.4\%$&$34.6\%$\\
phenotype&GA&GA&P&GA&P\\ \hline
$AKT$&$0$&$0$&$0$&$0$&$0$\\
$ATM_{lvl1}$&$0$&$0$&$0$&$0$&$0$\\
$ATM_{lvl2}$&$0$&$0$&$0$&$0$&$0$\\
$CDC25A$&$0$&$0$&$1$&$0$&$1$\\
$CHEK1\slash 2_{lvl1}$&$0$&$0$&$0$&$0$&$0$\\
$CHEK1\slash 2_{lvl2}$&$0$&$0$&$0$&$0$&$0$\\
$CyclinA$&$0$&$0$&$1$&$0$&$1$\\
$CyclinD1$&$0$&$0$&$0$&$0$&$1$\\
$CyclinE1$&$0$&$0$&$1$&$0$&$1$\\
$E2F1_{lvl1}$&$0$&$0$&$1$&$0$&$1$\\
$E2F1_{lvl2}$&$0$&$0$&$0$&$0$&$0$\\
$E2F3_{lvl1}$&$0$&$1$&$1$&$0$&$1$\\
$E2F3_{lvl2}$&$0$&$0$&$0$&$0$&$0$\\
$EGFR$&$0$&$0$&$0$&$0$&$0$\\
$FGFR3$&$1$&$1$&$1$&$1$&$1$\\
$GRB2$&$0$&$0$&$0$&$0$&$0$\\
$MDM2$&$0$&$0$&$0$&$0$&$0$\\
$p14ARF$&$0$&$0$&$1$&$0$&$0$\\
$p16INK4a$&$0$&$1$&$1$&$0$&$0$\\
$p21CIP$&$1$&$1$&$0$&$1$&$0$\\
$PI3K$&$0$&$0$&$0$&$0$&$0$\\
$PTEN$&$0$&$0$&$0$&$0$&$0$\\
$RAS$&$1$&$1$&$1$&$1$&$1$\\
$RB1$&$1$&$0$&$0$&$1$&$0$\\
$RBL2$&$1$&$1$&$0$&$1$&$0$\\
$SPRY$&$1$&$1$&$1$&$1$&$1$\\
$TP53$&$0$&$0$&$0$&$0$&$0$
\end{tabular}
\end{center}
\end{small}
where GA means growth arrest and P means proliferation.

The physiological variant is able to exhibit the two possible responses according to the input configuration: proliferation, represented by $a_{physio3}$, and growth arrest, represented by $a_{physio1}$ and $a_{physio2}$. Growth arrest occupies $85.2\%$ of the physiological state space, suggesting that normal cells are more likely to comply with growth inhibiting signals than with stimulating ones.

With the pathological variant modeling cells whose the two growth inhibitors p14ARF and p16INK4a are lost, the two possible responses are still present with again growth arrest being more likely than proliferation. Even if $a_{physio2}$ disappears, growth arrest is still possible with $a_{physio1}$ whose the basin increases from $10.7\%$ in $S_{physio}$ to $65.4\%$ in $S_{patho}$. The proliferating phenotype is also still possible but through the pathological attractor $a_{patho1}$ which, in a way, replaces the physiological attractor $a_{physio3}$.

However, the global tendency toward growth arrest significantly decreases: proliferation is more than twice likely in the pathological variant than in the physiological one with a shift from $14.8\%$ in $S_{physio}$ to $34.6\%$ in $S_{patho}$. Therefore, such pathological cells might be less responsive to growth inhibiting signals and more apt at proliferating, which is a major concern in tumorigenesis and consistent with the loss of two growth inhibitors.

To ensure that browsing the state space by performing $1\ 000$ random walks of $10\ 000$ steps is sufficient to find all the attractors while estimating their basin with little variability, the physiological and pathological attractor sets were computed $100$ times each:
\begin{small}
\begin{center}
\begin{tabular}{c|c|c}
set&attractor&basin ($\%$ of $S$)\\ \hline
\multirow{3}{*}{$A_{physio}$}&$a_{physio1}$&$10.518 \pm 0.833$\\
&$a_{physio2}$&$73.462 \pm 1.24$\\
&$a_{physio3}$&$16.02 \pm 1.091$\\ \hline
\multirow{2}{*}{$A_{patho}$}&$a_{physio1}$&$65.037 \pm 1.687$\\
&$a_{patho1}$&$34.963 \pm 1.687$
\end{tabular}
\end{center}
\end{small}

These results indicate that, in the present case study, browsing the state space by performing $1\ 000$ random walks of $10\ 000$ steps is enough robust to obtain reproducible results. Indeed, at each time, the same attractors are found: no attractor is missed. Moreover, the means of the basin estimations exhibit low standard deviations: basin estimations are subjected to variability but are nonetheless reliable.

\subsubsection{Therapeutic bullets}

As in the example network, bullets are assessed for their therapeutic potential on the pathological variant $\boldsymbol{f}_{patho}$ according to the new criterion: increasing the physiological part $\bigcup B_{physio,i}$ in the testing state space $S_{test}$ with a threshold of $5\%$. It means that therapeutic bullets have to push $\bigcup B_{physio,i}$ from $65.4\%$ in $S_{patho}$ to at least $65.4+5=70.4\%$ in $S_{test}$.

This case study belonging to a cancerous setting, it is desirable that therapeutic bullets also promote growth arrest in order to slow down tumorigenesis. In terms of basins and attractors, it means that interesting therapeutic bullets should decrease $B_{patho1}$, avoid $a_{physio3}$, increase $B_{physio1}$ and reintroduce $a_{physio2}$. Such therapeutic bullets could be qualified as anti-proliferative.

All the $1\ 458$ bullets made of one to two targets are tested. Among them, kali finds $9$ $1$-therapeutic bullets and $174$ $2$-therapeutic bullets listed in the supporting files \texttt{bladder\_B\_therap\_1.txt} and \texttt{bladder\_B\_therap\_2.txt} respectively. In addition to increasing the physiological part, all the returned therapeutic bullets are anti-proliferative. Indeed, all of them do not reintroduce $a_{physio3}$ and decrease $B_{patho1}$, thus promoting growth arrest through $a_{physio1}$ and\slash or $a_{physio2}$.

For example, the two following $1$-therapeutic bullets increase $B_{physio1}$ while decreasing $B_{patho1}$, thus exhibiting anti-proliferative effect as expected when targeting the well known growth promoting PI3K\slash Akt pathway \cite{mayer2016pi3k}:
\begin{small}
\begin{center}
\begin{tabular}{r|r c r|r r r r}
\multicolumn{1}{c}{bullet}&\multicolumn{3}{|c|}{gain}&$B_{physio1}$&$B_{physio2}$&$B_{physio3}$&$B_{patho1}$\\ \hline
$AKT[0]$&$65.4\%$&$\to$&$89.3\%$&$89.3\%$&$0\%$&$0\%$&$10.7\%$\\
$PI3K[0]$&$65.4\%$&$\to$&$86\%$&$86\%$&$0\%$&$0\%$&$14\%$
\end{tabular}
\end{center}
\end{small}
Below is an other interesting $1$-therapeutic bullet predicting that inhibiting CDC25A is anti-proliferative:
\begin{small}
\begin{center}
\begin{tabular}{r|r c r|r r r r}
\multicolumn{1}{c}{bullet}&\multicolumn{3}{|c|}{gain}&$B_{physio1}$&$B_{physio2}$&$B_{physio3}$&$B_{patho1}$\\ \hline
$CDC25A[0]$&$65.4\%$&$\to$&$100\%$&$100\%$&$0\%$&$0\%$&$0\%$
\end{tabular}
\end{center}
\end{small}

This therapeutic bullet is able to definitively suppress proliferation by making $B_{physio1}=100\%$ of $S_{test}$. It makes sense since the tyrosine phosphatase CDC25A can activate several cyclin-dependent kinases (CDKs) which, with their cyclin partners, promote cell cycle and then growth \cite{shen2012role}. This prediction correlates with biological knowledge about CDC25A inhibitors as potential anticancer agents \cite{lavecchia2009cdc25a}. For example, it is demonstrated that inhibiting CDC25A can suppress the growth of hepatocellular carcinoma cells \cite{xu2008cdc25a,kar2006pm}. Moreover, a recent work was specially dedicated to the synthesis of anticancer agents inhibiting the CDC25A\slash B phosphatases \cite{rostom2017structure}.

This highlights that dry-lab predictions consistent with factual evidences coming from wet-lab experiments are obtainable through kali, provided that the underlaying model is consistent too. Note that this does not imply that all the predictions are correct: needless to say that biological interpretation by experts is still mandatory.

The $2$-therapeutic bullets also bring some interesting predictions. For example, they indicate that sprouty (SPRY) could be a therapeutic target but only in combination with another one: there are no $1$-therapeutic bullets containing it. Sprouty negatively regulates mitogen-activated protein kinase (MAPK) signaling pathways downstream of growth factor receptors and is down-regulated in many cancers \cite{masoumi2014developing}. Consequently, stimulating sprouty should be anti-proliferative and this is what suggest the two following therapeutic bullets, even if the gain is relatively minor:
\begin{small}
\begin{center}
\begin{tabular}{r r|r c r|r r r}
\multicolumn{2}{c}{bullet}&\multicolumn{3}{|c|}{gain}&$B_{physio1}$&$B_{physio2}$&$B_{physio3}$\\
&&&&&$B_{patho1}$&&\\ \hline
$E2F3_{lvl2}[0]$&$SPRY[1]$&$65.4\%$&$\to$&$70.5\%$&$70.5\%$&$0\%$&$0\%$\\
&&&&&$29.5\%$&&\\ \hline
$MDM2[0]$&$SPRY[1]$&$65.4\%$&$\to$&$71.7\%$&$71.7\%$&$0\%$&$0\%$\\
&&&&&$28.3\%$&&
\end{tabular}
\end{center}
\end{small}

These two therapeutic bullets indicate that stimulating sprouty should be done along with an inhibition of MDM2 or E2F3. As with CDC25A, this prediction correlates with biological knowledge: MDM2 is a major inhibitor of the well known tumor suppressor p53 \cite{wang2017targeting} while E2F3 is a required transcription factor for the cell cycle \cite{leone1998e2f3}. However, only the level $2$ of E2F3 is concerned, meaning that only its over-expression should be prevented. In other words, this is not an inhibition of E2F3 but rather the prevention of its over-expression, if any.

In the returned therapeutic bullets there are also intriguing results such as the following one:
\begin{small}
\begin{center}
\begin{tabular}{r|r c r|r r r r}
\multicolumn{1}{c}{bullet}&\multicolumn{3}{|c|}{gain}&$B_{physio1}$&$B_{physio2}$&$B_{physio3}$&$B_{patho1}$\\ \hline
$FGFR3[1]$&$65.4\%$&$\to$&$74.1\%$&$74.1\%$&$0\%$&$0\%$&$25.9\%$
\end{tabular}
\end{center}
\end{small}
This therapeutic bullet moderately increases $B_{physio1}$ at the expense of $B_{patho1}$, thus promoting growth arrest. However, FGFR3 is a growth factor receptor and is frequently subjected to activating mutations in low grade bladder cancers \cite{billerey2001frequent}. Therefore, stimulating FGFR3 should promote proliferation, not growth arrest. However, Elisabeth Remy and colleagues have implemented a negative crosstalk from FGFR3 to the growth factor receptor EGFR in their model. This negative crosstalk may explain why stimulating FGFR3 is predicted to be anti-proliferative.

Indeed, $EGFR[0]$ is one of the returned therapeutic bullet and represent a direct inhibition of EGFR, a well studied target in cancer therapies \cite{seshacharyulu2012targeting,dhomen2012therapeutic}. Consequently and according to the model, $FGFR3[1]$ can be interpreted as an indirect inhibition of EGFR, especially since these two therapeutic bullets have almost identical effects in magnitude:
\begin{small}
\begin{center}
\begin{tabular}{r|r c r|r r r r}
\multicolumn{1}{c}{bullet}&\multicolumn{3}{|c|}{gain}&$B_{physio1}$&$B_{physio2}$&$B_{physio3}$&$B_{patho1}$\\ \hline
$EGFR[0]$&$65.4\%$&$\to$&$75.4\%$&$75.4\%$&$0\%$&$0\%$&$24.6\%$
\end{tabular}
\end{center}
\end{small}

Finally, it should be noted that the three following bullets are not predicted therapeutic by kali: $p14ARF[1]$, $p16INK4a[1]$ and $p14ARF[1]\ p16INK4a[1]$. As with the example network, this suggests that applying the opposite action of the pathological disturbance is not necessarily a therapeutic solution. Moreover, and again as with the example network, none of the found therapeutic bullets allow the recovery of all the physiological attractors: golden bullets seem to be as idealistic as golden pills.

\subsection{Computation times}

The results presented in this article were obtained on a laptop with $16$GB of RAM and an Intel Core i7-6600U processor. There are two kali parameters strongly influencing computation times. These two parameters control the attractor search and are:
\begin{itemize}
\item $max_{S}$: the maximum number of initial states to use when computing an attractor set
\item $max_{k}$: the length of the random walks performed to reach candidate attractors
\end{itemize}

The asynchronous attractor search consists in performing $max_{S}$ random walks of $max_{k}$ steps. Knowing that such a search is performed for computing an attractor set and that one attractor set is computed per tested bullet, the computation time can greatly increase with $max_{S}$ and\slash or $max_{k}$. By the way, computation times also increase with $n_{targ}$, $max_{targ}$ and $max_{moda}$, three kali parameters controlling how much bullets are tested:
\begin{itemize}
\item $n_{targ}$: the number of targets per bullet
\item $max_{targ}$: the maximum number of target combinations to test
\item $max_{moda}$: the maximum number of modality arrangements to test
\end{itemize}

The used logic can also increase computation times because the size of the state space is $h^{n}$, where $n$ is the number of nodes in the network and $h$ the number of possible values for the variables. For example, $h=2$ with Boolean logic and $h=3$ with $3$-valued logic. $h$ can also increase the number of testable bullets, and then computation times, since there are $(n! \cdot h^{n_{targ}})/(n_{targ}! \cdot (n-n_{targ})!)$ possible bullets.

Below are the computation times of the runs performed for this article:
\begin{small}
\begin{center}
\begin{tabular}{r|c|c|c}
&example network&example network&case study\\
&(Boolean)&($3$-valued)&(Boolean)\\ \hline
$max_{S}$&$512$ (all)&$1\ 000$&$1\ 000$\\
$max_{k}$&$1\ 000$&$1\ 000$&$10\ 000$\\
$1$-bullets&$18$ (all)&$27$ (all)&$54$ (all)\\
$2$-bullets&$144$ (all)&$324$ (all)&$1\ 404$ (all)\\ \hline
$A_{physio}$&$130$ms&$187$ms&$6$s$89$ms\\
$A_{patho}$&$109$ms&$218$ms&$6$s$55$ms\\
$B_{therap}$ ($n_{targ}=1$)&$2$s$510$ms&$6$s$775$ms&$5$m$57$s$950$ms\\
$B_{therap}$ ($n_{targ}=2$)&$19$s$133$ms&$1$m$23$s$526$ms&$2$h$43$m$36$s$709$ms
\end{tabular}
\end{center}
\end{small}

\section{Conclusion}

kali can now work on both synchronous and asynchronous Boolean networks. This is probably the most required improvement since asynchronous updating is frequently used in the scientific community and might be better realistic than synchrony, as discussed in the \hyperref[async]{\texttt{Introduction}} section page \pageref{async}. Consequently, a computational tool aimed at working on models built by the scientific community, such as kali, has to handle this updating scheme.

Also note that there are more than one asynchronous updating scheme. The one implemented in kali is the most popular and is named the general asynchronous updating: one randomly selected variable is updated at each iteration. However, other asynchronous updating methods exist. For example, with the random order updating, all the variables are updated at each iteration along a randomly selected order. Implementing various asynchronous updating schemes in kali could be a required future improvement.

kali now uses a new criterion for assessing therapeutic bullets. This new criterion brings a wider range of targeting strategies intended to push pathological behaviors toward physiological ones. It is based on a more permissive assumption stating that reducing the reachability of pathological attractors is therapeutic.

For an \textit{in silico} tool such as kali, being a little bit more permissive can be important since the findings obtained by simulations have to outlive the bottleneck separating predictions and reality. With a too strict assessment of therapeutic bullets, the risk of highlighting too few candidate targets or to miss some interesting ones can be high. Moreover, predicted does not necessarily mean true: a prediction of apparently poor interest can reveal itself of great interest, and \textit{vice versa}.

This new criterion also brings a finer assessment of therapeutic bullets since all the possible increases of $\bigcup B_{physio,i}$ in $S_{test}$ are considered. With the previous criterion, there was only one therapeutic potential: $\bigcup B_{physio,i}=100\%$ of $S_{test}$, thus reducing the assessment of bullets to therapeutic or not. Things are not so dichotomous but rather nuanced: the assessment of therapeutic bullets should be nuanced too.

kali can now work with multivalued logic. Allowing variables to take an arbitrary finite number of values should enable to more accurately model biological processes and produce more fine-tuned therapeutic bullets. However, this accuracy and fine-tuning are at the cost of an increased computational requirement. Indeed, the size of the state space depends on the size of the model and the used logic.

Consequently, the size of the model and the used logic should be balanced: the smaller the model is, the more variables should be finely valued. For example, for an accurate therapeutic investigation, the model should only contain the essential and specific pieces of the studied pathological mechanisms modeled by a finely valued logic. On the other hand, for a broad therapeutic investigation, a more exhaustive model can be used but modeled by a coarse-grained logic.

Note that the ultimate multivalued logic is the infinitely valued one, which is fuzzy logic \cite{zadeh1988fuzzy}. With fuzzy logic, the whole interval of real numbers $[0;1]$ is used to valuate variables, which might bring the best accuracy for the qualitative modeling formalism \cite{poret2014enhancing,morris2011training,aldridge2009fuzzy}. However, using such a continuous logic implies to leave the relatively convenient discrete paradigm to enter the continuous one where, for example, the state space is infinite.

kali also demonstrates that it is able at predicting therapeutic bullets consistent with the underlaying model, with biological knowledge and with experimental evidences. For example, in the bladder tumorigensis case study, kali returned therapeutic bullets inhibiting the PI3K\slash Akt pathway or the CDC25A tyrosine phosphatase, two documented targets in cancer therapies.

Even the surprising $FGFR3[1]$ therapeutic bullets, which suggests to stimulate a growth factor receptor for promoting growth arrest, is consistent with the underlaying model. Indeed, according to this model, it appears that $FGFR3[1]$ is founded in a negative crosstalk from FGFR3 to EGFR, thus indirectly inhibiting the growth factor receptor EGFR, which is also a documented target in cancer therapies.

Two additional improvements are envisioned for kali. The first one is to allow therapeutic bullets to create new attractors, namely \textit{de novo} attractors. It is conceivable that a bullet can greatly decrease pathological basins while creating a new attractor not belonging to the physiological variant nor to the pathological one. Such a \textit{de novo} attractor is currently tagged by kali as not physiological and then pathological, thus rejecting the concerned bullet. However, if a \textit{de novo} attractor is weakly pathological and induced by a bullet greatly decreasing the basin of other and heavier pathological attractors, such a case should be retained.

The second envisioned improvement is to allow partial matching when checking if an attractor is associated with a physiological phenotype by comparing it to the physiological attractors. Currently, an attractor which does not match a physiological attractor is considered pathologic. However, it is conceivable that some variables not exhibiting a physiological behavior in an attractor do not pathologically impact its associated phenotype. To allow such a case to be considered, some variables within attractors should be allowed to not be matched when assessing the associated phenotype.

This suggests the concept of decisive variables, namely variables whose the behavior in the attractors is sufficient to biologically interpret the associated phenotypes. Elisabeth Remy and colleagues have already implemented this distinction in their model of bladder tumorigenesis used in the present article as case study: decisive variables are those belonging to the equations of the three output phenotypes. Therefore, kali could allow non-decisive variables to not be matched.

Ultimately, this could allow the modeler to specify himself\slash herself what a physiological attractor is without having to consider a physiological and a pathological variant. This could also allow to no longer think in terms of physiological \textit{versus} pathological attractors but just desirable ones. Moreover, implementing the second envisioned improvement could greatly facilitate the implementation of the first one since the goal would become to obtain desired attractors regardless if they are \textit{de novo} or not.

\newpage

\section{Appendix 1: recall of previous concepts}

\label{brief} Below are some important concepts introduced in the previous article where the complete background was presented \cite{poret2014silico}.

\subsection{Biological networks}

A network is a directed graph $G=(V,E)$ where $V=\lbrace v_{1},\ldots,v_{n}\rbrace$ is the set containing the nodes of the network and $E=\lbrace (v_{i,1},v_{j,1}),\ldots,(v_{i,m},v_{j,m})\rbrace$ is the set containing the edges linking these nodes. In practice, nodes represent entities while edges represent binary relations $R \subset V^{2}$ involving them: $v_{i}\ R\ v_{j}$ \cite{zhu2007getting}. It indicates that the node $v_{i}$ exerts an influence on the node $v_{j}$. For example, in gene regulatory networks \cite{emmert2014gene}, $v_{i}$ can be a transcription factor while $v_{j}$ a gene product. The edges are frequently signed so that they indicate if $v_{i}$ exerts a positive or a negative influence on $v_{j}$, such as an activation or an inhibition.

\subsection{Boolean networks}

A Boolean network is a network where nodes are Boolean variables $x_{i}$ and edges $(x_{i},x_{j})$ are the $is\ input\ of$ relation: $x_{i}\ is\ input\ of\ x_{j}$. Each variable $x_{i}$ has $b_{i} \in [\![0,n]\!]$ inputs influencing its state. Note that $b_{i}=0$ is possible. In this case, $x_{i}$ is an input of the network. Depending on the updating scheme, at each iteration $k \in [\![k_{0},k_{end}]\!]$ one or more $x_{i}$ are updated using their associated Boolean transition function $f_{i}$. This function uses Boolean operators, typically $\land$ ($and$), $\lor$ ($or$) and $\lnot$ ($not$), to specify how the inputs $x_{i,1},\ldots,x_{i,b_{i}}$ of $x_{i}$ have to be related to compute its value, as in the following pseudocode representing a synchronous updating:\\

\noindent \textbf{for} $k \gets k_{0},\ldots,k_{end}$\\
\indent $x_{1} \gets f_{1}(x_{1,1},\ldots,x_{1,b_{1}})$\\
\indent $\vdots$\\
\indent $x_{n} \gets f_{n}(x_{n,1},\ldots,x_{n,b_{n}})$\\
\textbf{end for}\\

\noindent which can be written in a more concise form:\\

\noindent \textbf{for} $k \gets k_{0},\ldots,k_{end}$\\
\indent $\boldsymbol{x} \gets \boldsymbol{f}(\boldsymbol{x})$\\
\textbf{end for}\\

\noindent where $\boldsymbol{f}=(f_{1},\ldots,f_{n})$ is the Boolean transition function of the network and $\boldsymbol{x}=(x_{1},\ldots,x_{n})$ is its state vector. The value of the state vector belongs to the state space $S=\lbrace 0,1\rbrace^{n}$, which is the set containing all the possible states of the network.

The set $A=\lbrace a_{1},\ldots,a_{p}\rbrace$ containing the attractors of the network is its attractor set. An attractor $a_{i}$ is a collection of states $(\boldsymbol{x}_{1},\ldots,\boldsymbol{x}_{q})$ such that once the system reaches a state $\boldsymbol{x}_{j} \in a_{i}$, it can subsequently visit the states of $a_{i}$ but no other ones: the system can not escape. The set $B_{i} \subset S$ containing the states $\boldsymbol{x} \in S$ from which $a_{i}$ can be reached is its basin of attraction, or simply basin.

\subsection{Definitions}

\begin{itemize}
\item \textbf{physiological phenotype}: a phenotype which does not impair the life quantity\slash quality of the organism which exhibits it
\item \textbf{pathological phenotype}: a phenotype which impairs the life quantity\slash quality of the organism which exhibits it
\item \textbf{variant (of a biological network)}: given a biological network, a variant is one of its versions, namely the network plus eventually some modifications
\item \textbf{physiological variant}: a variant which produces only physiological phenotypes, this is the biological network as it should be, the one of healthy organisms
\item \textbf{pathological variant}: a variant which produces at least one pathological phenotype, this is a dysfunctional version of the biological network, a version found in ill organisms
\item \textbf{physiological attractor set}: the attractor set $A_{physio}$ of the physiological variant
\item \textbf{pathological attractor set}: the attractor set $A_{patho}$ of the pathological variant
\item \textbf{physiological Boolean transition function}: the Boolean transition function $\boldsymbol{f}_{physio}$ of the physiological variant
\item \textbf{pathological Boolean transition function}: the Boolean transition function $\boldsymbol{f}_{patho}$ of the pathological variant
\item \textbf{physiological attractor}: an attractor $a_{i}$ such that $a_{i} \in A_{physio}$, note that it does not exclude the possibility that $a_{i} \in A_{patho}$ in addition to $a_{i} \in A_{physio}$
\item \textbf{pathological attractor}: an attractor $a_{i}$ such that $a_{i} \notin A_{physio}$
\item \textbf{modality}: the perturbation $moda_{i} \in \lbrace 0,1 \rbrace$ applied on a node $v_{j}$ of the network, either activating ($moda_{i}=1$) or inactivating ($moda_{i}=0$), at each iteration $moda_{i}$ overwrites $f_{j}(\boldsymbol{x})$ making $x_{j}=moda_{i}$
\item \textbf{target}: a node $targ_{i}$ of the network on which a modality $moda_{i}$ is applied
\item \textbf{bullet}: a couple $(c_{targ},c_{moda})$ where $c_{targ}=(targ_{1},\ldots,targ_{r})$ is a combination without repetition of $r$ targets and $c_{moda}=(moda_{1},\ldots,moda_{r})$ is an arrangement with repetition of $r$ modalities, $moda_{i}$ is intended to be applied on $targ_{i}$
\end{itemize}

\newpage

\section{Appendix 2: multivalued case}

\label{multivalued} Below is the multivalued version of the example network:
\begin{small}
\begin{IEEEeqnarray*}{r C l}
do&=&do\\
factory&=&factory\\
energy&=&max(min(energy,1-task),factory)\\
locker&=&1-energy\\
releaser&=&do\\
sequester&=&1-releaser\\
activator&=&min(do,1-locker)\\
effector&=&min(activator,1-sequester)\\
task&=&effector
\end{IEEEeqnarray*}
\end{small}
where the Boolean operators are replaced by the Zadeh ones.

To take advantage of multivalued logic, $f_{locker}$ becomes $locker=min(1-energy,0.5)$ in $\boldsymbol{f}_{patho}$. This equation tells that the locker is actionable when required, namely when there is no energy, but that it is unable at being fully operational due to some pathological defects: the maximal value of $f_{locker}$ in $\boldsymbol{f}_{patho}$ is $0.5$.

As mentioned in the article, $0.5$ can be interpreted as an incomplete activation\slash inhibition depending on what is modeled. Consequently, in the pathological variant, the activator is at most partly inhibited by the locker when no energy is available, allowing the task to be nevertheless performed. However, in this case, the task is itself moderately performed.

\subsection{Attractor sets}

The example network is computed asynchronously using a $3$-valued logic. To compute an attractor set, kali performs $1\ 000$ random walks of $1\ 000$ steps. Below are the returned attractors:
\begin{itemize}
\item $A_{physio}$:
\begin{small}
\begin{center}
\begin{tabular}{c|c|c|c|c|c|c}
attractor&basin ($\%$ of $S_{physio}$)&$do$&$factory$&$energy$&$locker$&$task$\\ \hline
$a_{physio1}$&$6.1\%$&$0$&$0$&$0$&$1$&$0$\\
$a_{physio2}$&$4.5\%$&$0$&$0$&$0.5$&$0.5$&$0$\\
$a_{physio3}$&$2.5\%$&$0$&$0$&$1$&$0$&$0$\\
$a_{physio4}$&$9.7\%$&$0$&$0.5$&$0.5$&$0.5$&$0$\\
$a_{physio5}$&$1.8\%$&$0$&$0.5$&$1$&$0$&$0$\\
$a_{physio6}$&$10.8\%$&$0$&$1$&$1$&$0$&$0$\\
$a_{physio7}$&$6.5\%$&$0.5$&$0$&$0$&$1$&$0$\\
$a_{physio8}$&$4.8\%$&$0.5$&$0$&$0.5$&$0.5$&$0.5$\\
$a_{physio9}$&$10.3\%$&$0.5$&$0.5$&$0.5$&$0.5$&$0.5$\\
$a_{physio10}$&$10.6\%$&$0.5$&$1$&$1$&$0$&$0.5$\\
$a_{physio11}$&$7.3\%$&$1$&$0$&$0$&$1$&$0$\\
$a_{physio12}$&$3.2\%$&$1$&$0$&$0.5$&$0.5$&$0.5$\\
$a_{physio13}$&$10.3\%$&$1$&$0.5$&$0.5$&$0.5$&$0.5$\\
$a_{physio14}$&$11.6\%$&$1$&$1$&$1$&$0$&$1$
\end{tabular}
\end{center}
\end{small}
\item $A_{patho}$:
\begin{small}
\begin{center}
\begin{tabular}{c|c|c|c|c|c|c}
attractor&basin ($\%$ of $S_{patho}$)&$do$&$factory$&$energy$&$locker$&$task$\\ \hline
$a_{patho1}$&$6.2\%$&$0$&$0$&$0$&$0.5$&$0$\\
$a_{physio2}$&$4.7\%$&$0$&$0$&$0.5$&$0.5$&$0$\\
$a_{physio3}$&$2.2\%$&$0$&$0$&$1$&$0$&$0$\\
$a_{physio4}$&$9.7\%$&$0$&$0.5$&$0.5$&$0.5$&$0$\\
$a_{physio5}$&$1.8\%$&$0$&$0.5$&$1$&$0$&$0$\\
$a_{physio6}$&$10.8\%$&$0$&$1$&$1$&$0$&$0$\\
$a_{patho2}$&$5.5\%$&$0.5$&$0$&$0$&$0.5$&$0.5$\\
$a_{physio8}$&$5.8\%$&$0.5$&$0$&$0.5$&$0.5$&$0.5$\\
$a_{physio9}$&$10.3\%$&$0.5$&$0.5$&$0.5$&$0.5$&$0.5$\\
$a_{physio10}$&$10.6\%$&$0.5$&$1$&$1$&$0$&$0.5$\\
$a_{patho3}$&$7.3\%$&$1$&$0$&$0$&$0.5$&$0.5$\\
$a_{physio12}$&$3.2\%$&$1$&$0$&$0.5$&$0.5$&$0.5$\\
$a_{physio13}$&$10.3\%$&$1$&$0.5$&$0.5$&$0.5$&$0.5$\\
$a_{physio14}$&$11.6\%$&$1$&$1$&$1$&$0$&$1$
\end{tabular}
\end{center}
\end{small}
\end{itemize}

$a_{physio1}$, $a_{physio3}$, $a_{physio6}$, $a_{physio11}$ and $a_{physio14}$ are the physiological attractors found in the Boolean case with a different numbering due to additional attractors coming from multivalued logic. Indeed, given that $\lbrace 0,1 \rbrace \subset \lbrace 0,0.5,1 \rbrace$ and that the Zadeh operators also work with Boolean logic, the Boolean results are still obtainable. The same does not apply to the pathological attractors because $f_{locker}$ in $\boldsymbol{f}_{patho}$ differs between the Boolean and multivalued cases.

For example, $a_{physio13}$ indicates that the do instruction is sent while energy is partly supplied. Consequently, the locker is partly activated resulting in a partial inhibition of the activator. The task is thus moderately performed despite full do instruction, hence coping with moderate energy supply.

Concerning the pathological attractors, as an example $a_{patho3}$ indicates that the do instruction is sent in absence of energy supply. Consequently, the locker should be fully activated to prevent the task. However, due to some pathological defects, the locker is at most partly activated. The task is then performed in absence of energy. However, since the locker is partly operational, the task is not performed at its maximum rate, maybe limiting pathological consequences.

Among the pathological attractors, $a_{patho1}$ can be considered weakly pathological. Indeed, in $a_{patho1}$ the locker should be fully activated since there is no energy. However, there is no do instruction and therefore no task to stop. On the other hand, $a_{patho2}$ and $a_{patho3}$ are more pathological since the task is performed while no energy is available.

\subsection{Therapeutic bullets}

All the bullets made of one to two targets are tested with a threshold of $5\%$. Below are the returned therapeutic bullets:

\begin{landscape}

\begin{itemize}
\item $1$-therapeutic bullets:
\begin{tiny}
\begin{center}
\begin{tabular}{r|r c r|r r r r r r r r r}
\multicolumn{1}{c}{bullet}&\multicolumn{3}{|c|}{gain}&$B_{physio1}$&$B_{physio2}$&$B_{physio3}$&$B_{physio4}$&$B_{physio5}$&$B_{physio6}$&$B_{physio7}$&$B_{physio8}$&$B_{physio9}$\\
&&&&$B_{physio10}$&$B_{physio11}$&$B_{physio12}$&$B_{physio13}$&$B_{physio14}$&$B_{patho1}$&$B_{patho2}$&$B_{patho3}$&\\ \hline
$factory[0.5]$&$81\%$&$\to$&$100\%$&$0\%$&$0\%$&$0\%$&$29.3\%$&$6.1\%$&$0\%$&$0\%$&$0\%$&$32.2\%$\\
&&&&$0\%$&$0\%$&$0\%$&$32.4\%$&$0\%$&$0\%$&$0\%$&$0\%$&\\ \hline
$factory[1]$&$81\%$&$\to$&$100\%$&$0\%$&$0\%$&$0\%$&$0\%$&$0\%$&$35.4\%$&$0\%$&$0\%$&$0\%$\\
&&&&$32.2\%$&$0\%$&$0\%$&$0\%$&$32.4\%$&$0\%$&$0\%$&$0\%$&
\end{tabular}
\end{center}
\end{tiny}
\item $2$-therapeutic bullets:
\begin{tiny}
\begin{center}
\begin{tabular}{r r|r c r|r r r r r r r r r}
\multicolumn{2}{c}{bullet}&\multicolumn{3}{|c|}{gain}&$B_{physio1}$&$B_{physio2}$&$B_{physio3}$&$B_{physio4}$&$B_{physio5}$&$B_{physio6}$&$B_{physio7}$&$B_{physio8}$&$B_{physio9}$\\
&&&&&$B_{physio10}$&$B_{physio11}$&$B_{physio12}$&$B_{physio13}$&$B_{physio14}$&$B_{patho1}$&$B_{patho2}$&$B_{patho3}$&\\ \hline
$do[0]$&$factory[0.5]$&$81\%$&$\to$&$100\%$&$0\%$&$0\%$&$0\%$&$84\%$&$16\%$&$0\%$&$0\%$&$0\%$&$0\%$\\
&&&&&$0\%$&$0\%$&$0\%$&$0\%$&$0\%$&$0\%$&$0\%$&$0\%$&\\ \hline
$do[0]$&$factory[1]$&$81\%$&$\to$&$100\%$&$0\%$&$0\%$&$0\%$&$0\%$&$0\%$&$100\%$&$0\%$&$0\%$&$0\%$\\
&&&&&$0\%$&$0\%$&$0\%$&$0\%$&$0\%$&$0\%$&$0\%$&$0\%$&\\ \hline
$do[0.5]$&$factory[0.5]$&$81\%$&$\to$&$100\%$&$0\%$&$0\%$&$0\%$&$0\%$&$0\%$&$0\%$&$0\%$&$0\%$&$100\%$\\
&&&&&$0\%$&$0\%$&$0\%$&$0\%$&$0\%$&$0\%$&$0\%$&$0\%$&\\ \hline
$do[0.5]$&$factory[1]$&$81\%$&$\to$&$100\%$&$0\%$&$0\%$&$0\%$&$0\%$&$0\%$&$0\%$&$0\%$&$0\%$&$0\%$\\
&&&&&$100\%$&$0\%$&$0\%$&$0\%$&$0\%$&$0\%$&$0\%$&$0\%$&\\ \hline
$do[1]$&$factory[0.5]$&$81\%$&$\to$&$100\%$&$0\%$&$0\%$&$0\%$&$0\%$&$0\%$&$0\%$&$0\%$&$0\%$&$0\%$\\
&&&&&$0\%$&$0\%$&$0\%$&$100\%$&$0\%$&$0\%$&$0\%$&$0\%$&\\ \hline
$do[1]$&$factory[1]$&$81\%$&$\to$&$100\%$&$0\%$&$0\%$&$0\%$&$0\%$&$0\%$&$0\%$&$0\%$&$0\%$&$0\%$\\
&&&&&$0\%$&$0\%$&$0\%$&$0\%$&$100\%$&$0\%$&$0\%$&$0\%$&\\ \hline
$do[0]$&$energy[1]$&$81\%$&$\to$&$100\%$&$0\%$&$0\%$&$34.9\%$&$0\%$&$32.1\%$&$33\%$&$0\%$&$0\%$&$0\%$\\
&&&&&$0\%$&$0\%$&$0\%$&$0\%$&$0\%$&$0\%$&$0\%$&$0\%$&\\ \hline
$do[0]$&$task[0]$&$81\%$&$\to$&$89\%$&$0\%$&$11.6\%$&$12.3\%$&$21.3\%$&$10.8\%$&$33\%$&$0\%$&$0\%$&$0\%$\\
&&&&&$0\%$&$0\%$&$0\%$&$0\%$&$0\%$&$11\%$&$0\%$&$0\%$&\\ \hline
$do[0.5]$&$task[0.5]$&$81\%$&$\to$&$89.4\%$&$0\%$&$0\%$&$0\%$&$0\%$&$0\%$&$0\%$&$0\%$&$24.3\%$&$32.1\%$\\
&&&&&$33\%$&$0\%$&$0\%$&$0\%$&$0\%$&$0\%$&$10.6\%$&$0\%$&\\ \hline
$factory[0]$&$energy[0.5]$&$81\%$&$\to$&$100\%$&$0\%$&$35.4\%$&$0\%$&$0\%$&$0\%$&$0\%$&$0\%$&$32.2\%$&$0\%$\\
&&&&&$0\%$&$0\%$&$32.4\%$&$0\%$&$0\%$&$0\%$&$0\%$&$0\%$&\\ \hline
$factory[0.5]$&$energy[0.5]$&$81\%$&$\to$&$100\%$&$0\%$&$0\%$&$0\%$&$35.4\%$&$0\%$&$0\%$&$0\%$&$0\%$&$32.2\%$\\
&&&&&$0\%$&$0\%$&$0\%$&$32.4\%$&$0\%$&$0\%$&$0\%$&$0\%$&\\ \hline
$factory[1]$&$energy[1]$&$81\%$&$\to$&$100\%$&$0\%$&$0\%$&$0\%$&$0\%$&$0\%$&$35.4\%$&$0\%$&$0\%$&$0\%$\\
&&&&&$32.2\%$&$0\%$&$0\%$&$0\%$&$32.4\%$&$0\%$&$0\%$&$0\%$&\\ \hline
$factory[1]$&$locker[0]$&$81\%$&$\to$&$100\%$&$0\%$&$0\%$&$0\%$&$0\%$&$0\%$&$35.4\%$&$0\%$&$0\%$&$0\%$\\
&&&&&$32.2\%$&$0\%$&$0\%$&$0\%$&$32.4\%$&$0\%$&$0\%$&$0\%$&
\end{tabular}
\end{center}
\end{tiny}
\end{itemize}

\end{landscape}

For example, the therapeutic bullet $factory[1]\ locker[0]$ is interesting. It suppresses all the pathological attractors while maintaining three physiological attractors allowing the pathological variant to properly respond to the three possible levels of the do instruction. Moreover, the basins of these three physiological attractors, namely $a_{physio6}$, $a_{physio10}$ and $a_{physio14}$, equally span the state space, making them equally reachable.

On the other hand, the therapeutic bullet $do[0.5]\ factory[0.5]$ seems to be less interesting. While this bullet also suppresses all the pathological attractors, it enables only one physiological attractor. In this physiological attractor, namely $a_{physio9}$, all the variables are at their intermediate level: the network can not fulfill its switching function.

\newpage

\section{Appendix 3: core of kali}

\label{pseudocode} Below is the core of kali in pseudocode derived from its Go \cite{Go} sources, freely available on GitHub at \url{https://github.com/arnaudporet/kali} under the GNU General Public License \cite{GPL}. Note that the code may have evolved since the publication of the present article.

\subsection{Defined types}

\noindent \textbf{structure} $Attractor$\textcolor{gray}{// an attractor}\\
\indent \textbf{field} $Name$\textcolor{gray}{// its name, either $a_{physio}$ or $a_{patho}$}\\
\indent \textbf{field} $Basin$\textcolor{gray}{// the size of its basin, in percents of the state space}\\
\indent \textbf{field} $States$\textcolor{gray}{// its states, as a matrix of one state per row}\\
\textbf{end structure}\\

\noindent \textbf{structure} $Bullet$\textcolor{gray}{// a bullet}\\
\indent \textbf{field} $Targ$\textcolor{gray}{// its target combination, as a vector}\\
\indent \textbf{field} $Moda$\textcolor{gray}{// its modality arrangement, as a vector}\\
\indent \textbf{field} $Gain$\textcolor{gray}{// its gain, see below}\\
\indent \textbf{field} $Cover$\textcolor{gray}{// the size of each basin under its influence, see below}\\
\textbf{end structure}\\

\noindent $b.Gain$ is a vector $(gain_{1},gain_{2})$ where:
\begin{itemize}
\item $gain_{1}$ is the size of $\bigcup B_{physio,i}$ in $S_{patho}$
\item $gain_{2}$ is the size of $\bigcup B_{physio,i}$ in $S_{test}$
\end{itemize}
in $\%$ of $S_{patho}$ and $\%$ of $S_{test}$ respectively.\\

\noindent $b.Cover$ is a vector containing the size of the physiological and pathological basins in the testing state space, in percents of it.\\

\subsection{Parameters}

\noindent $nodes$\textcolor{gray}{// the node names, as a vector}\\
$\Omega$\textcolor{gray}{// the domain of the used logic, as a vector}\\
$sync$\textcolor{gray}{// use synchronous updating ($sync=1$) or not ($sync=0$)}\\
$whole$\textcolor{gray}{// build the whole state space ($whole=1$) or not ($whole=0$)}\\
$max_{S}$\textcolor{gray}{// the maximal size of the state space sample when $whole=0$}\\
$max_{k}$\textcolor{gray}{// the number of steps for the random walks (asynchronous only)}\\
$n_{targ}$\textcolor{gray}{// the number of targets per bullet}\\
$max_{targ}$\textcolor{gray}{// the maximum number of target combinations to test}\\
$max_{moda}$\textcolor{gray}{// the maximum number of modality arrangements to test}\\
$\delta$\textcolor{gray}{// the threshold for a bullet to be therapeutic, in percents of the state space}\\

\noindent When $whole=0$, a subset of the state space is built. This subset contains the initial states from which trajectories are performed. These trajectories are used for computing an attractor set, that is to reach the attractors. Note that these trajectories are free to evolve in all the state space. In other words, kali does not run on a subset of the state space, these are the initial states which are a subset of the state space.\\

\noindent To be considered therapeutic, a bullet has to make $gain_{2}-gain_{1} \geq \delta$ while not creating \textit{de novo} attractors.\\

\subsection{Functions}

\noindent \textbf{function} $DoTheJob(f_{physio},f_{patho},n_{targ},max_{targ},max_{moda},max_{S},max_{k},\delta,\\ \indent sync,nodes,\Omega,whole)$\\
\indent \textcolor{gray}{// do the job, this is the main function}\\
\indent $n \gets Size(nodes)$\textcolor{gray}{// the dimension of the state space $S$}\\
\indent \textbf{select} $whole$\\
\indent \indent \textbf{case} $0$\textcolor{gray}{// build a sample of $S$}\\
\indent \indent \indent $S \gets GenArrangs(\Omega,n,max_{S})$\\
\indent \indent \textbf{case} $1$\textcolor{gray}{// build all $S$}\\
\indent \indent \indent $S \gets GenSpace(\Omega,n)$\\
\indent \textbf{end select}\\
\indent $A_{physio} \gets ComputeAttractorSet(f_{physio},S,\varnothing,max_{k},0,sync,\emptyset)$\\
\indent $A_{patho} \gets ComputeAttractorSet(f_{patho},S,\varnothing,max_{k},1,sync,A_{physio})$\\
\indent $A_{versus} \gets GetVersus(A_{patho})$\textcolor{gray}{// the pathological attractors, see below}\\
\indent $C_{targ} \gets GenCombis(\lbrace 1,\ldots,n\rbrace,n_{targ},max_{targ})$\textcolor{gray}{// the target combinations}\\
\indent $C_{moda} \gets GenArrangs(\Omega,n_{targ},max_{moda})$\textcolor{gray}{// the modality arrangements}\\
\indent \textbf{if} $A_{versus} \neq \emptyset$\textcolor{gray}{// there are pathological basins to shrink}\\
\indent \indent $B_{therap} \gets ComputeTherapeuticBullets(f_{patho},S,C_{targ},C_{moda},max_{k},\\ \indent \indent \delta,sync,A_{physio},A_{patho},A_{versus})$\textcolor{gray}{// therapeutic bullets}\\
\indent \textbf{end if}\\
\indent \textbf{return} $S,A_{physio},A_{patho},A_{versus},C_{targ},C_{moda},B_{therap}$\\
\textbf{end function}\\

\noindent $Size(container)$ returns the number of items in $container$.\\

\noindent $GenSpace(\Omega,n)$ returns the $n$-dimensional state space of the vectors made of $n$ values from $\Omega$, as a matrix of one state vector per row.\\

\noindent $GenArrangs(\Omega,n,max_{arrang})$ returns $max_{arrang}$ arrangements with repetition made of $n$ elements from $\Omega$, as a matrix of one arrangement per row. If $max_{arrang}$ exceeds its maximal possible value then it is automatically decreased to its maximal possible value.\\

\noindent $GenCombis(\Omega,n,max_{combi})$ returns $max_{combi}$ combinations without repetition made of $n$ elements from $\Omega$, as a matrix of one combination per row. If $max_{combi}$ exceeds its maximal possible value then it is automatically decreased to its maximal possible value.\\

\noindent As explained later, the function $ComputeAttractorSet$ can use an already computed attractor set, namely the reference set, to name the attractors.\\

\noindent $A_{physio}$ is computed without bullet ($b \gets \varnothing$), without reference set ($A_{ref} \gets \emptyset$) and with the physiological setting ($setting \gets 0$).\\

\noindent $A_{patho}$ is computed without bullet ($b \gets \varnothing$), with a reference set ($A_{ref} \gets A_{physio}$) and with the pathological setting ($setting \gets 1$).\\

\noindent $A_{versus}$ is not a true attractor set but the set containing the pathological attractors: $A_{versus} \subset A_{patho}$. $A_{patho}$ can contains physiological attractors if the pathological variant exhibits some of them. However, $A_{versus}$ only contains the pathological attractors.\\

\noindent Therapeutic bullets are computed only if there are pathological basins to shrink, namely only if $A_{versus} \neq \emptyset$.\\

\noindent Note that target combinations are combinations of positions in the state vector: targets are identified by their position in the state vector, not by their name.\\

\noindent \textbf{function} $f_{physio}(x)$\\
\indent \textcolor{gray}{// update the state vector of the physiological variant}\\
\indent $y[1] \gets f_{physio}[1](x)$\textcolor{gray}{// update $x_{1}$ with $f_{physio,1}$}\\
\indent $\vdots$\\
\indent $y[n] \gets f_{physio}[n](x)$\textcolor{gray}{// update $x_{n}$ with $f_{physio,n}$}\\
\indent \textbf{return} $y$\textcolor{gray}{// the updated state vector}\\
\textbf{end function}\\

\noindent \textbf{function} $f_{patho}(x)$\\
\indent \textcolor{gray}{// update the state vector of the pathological variant}\\
\indent $y[1] \gets f_{patho}[1](x)$\textcolor{gray}{// update $x_{1}$ with $f_{patho,1}$}\\
\indent $\vdots$\\
\indent $y[n] \gets f_{patho}[n](x)$\textcolor{gray}{// update $x_{n}$ with $f_{patho,n}$}\\
\indent \textbf{return} $y$\textcolor{gray}{// the updated state vector}\\
\textbf{end function}\\

\noindent \textbf{function} $ComputeAttractor(f,x_{0},b,max_{k},sync)$\\
\indent \textcolor{gray}{// from $x_{0}$, reach an attractor $a$}\\
\indent \textbf{select} $sync$\\
\indent \indent \textbf{case} $1$\textcolor{gray}{// search a cycle}\\
\indent \indent \indent $a.States \gets ReachCycle(f,x_{0},b)$\\
\indent \indent \textbf{case} $0$\textcolor{gray}{// search a terminal SCC}\\
\indent \indent \indent \textbf{for}\\
\indent \indent \indent \indent $a.States \gets GoForward(f,Walk(f,x_{0},b,max_{k}),b)$\textcolor{gray}{// a candidate}\\
\indent \indent \indent \indent \textbf{if} $IsTerminal(a,f,b)$\textcolor{gray}{// the candidate attractor is a terminal SCC}\\
\indent \indent \indent \indent \indent \textbf{break}\textcolor{gray}{// then it is an asynchronous attractor}\\
\indent \indent \indent \indent \textbf{end if}\\
\indent \indent \indent \textbf{end for}\\
\indent \textbf{end select}\\
\indent \textbf{return} $a$\\
\textbf{end function}\\

\noindent If $ComputeAttractor(f,x_{0},b,max_{k},sync)$ is run with $sync=0$ (i.e. asynchronous case) then ensure that $max_{k}$ is big enough for random walks to reach attractors with high probability. If $max_{k}$ is too small and if there is no attractor near $x_{0}$, then this function will run indefinitely since it loops until an attractor is found starting from $x_{0}$. The default value of $max_{k}$ should be $10\ 000$. It can be smaller for little networks and should be higher for large networks.\\

\noindent \textbf{function} $ComputeAttractorSet(f,S,b,max_{k},setting,sync,A_{ref})$\\
\indent \textcolor{gray}{// compute an attractor set $A$, namely $A_{physio}$, $A_{patho}$ or $A_{test}$}\\
\indent $A \gets \lbrace \rbrace$\\
\indent \textbf{select} $setting$\textcolor{gray}{// select the default name for attractors}\\
\indent \indent \textbf{case} $0$\textcolor{gray}{// physiological setting}\\
\indent \indent \indent $name \gets a_{physio}$\\
\indent \indent \textbf{case} $1$\textcolor{gray}{// pathological setting}\\
\indent \indent \indent $name \gets a_{patho}$\\
\indent \textbf{end select}\\
\indent \textbf{for} $i \gets 1,\ldots,Size(S)$\textcolor{gray}{// browse $S$}\\
\indent \indent $a \gets ComputeAttractor(f,S[i],b,max_{k},sync)$\\
\indent \indent \textbf{if} $\exists i_{A}: A[i_{A}]=a$\textcolor{gray}{// $a$ is already found}\\
\indent \indent \indent $A[i_{A}].Basin \gets A[i_{A}].Basin+1$\textcolor{gray}{// then increase its basin}\\
\indent \indent \textbf{else}\textcolor{gray}{// new attractor}\\
\indent \indent \indent $a.Basin \gets 1$\textcolor{gray}{// then begin its basin}\\
\indent \indent \indent $A \gets A \cup \lbrace a\rbrace$\textcolor{gray}{// and add it to the attractor set}\\
\indent \indent \textbf{end if}\\
\indent \textbf{end for}\\
\indent \textbf{for} $i \gets 1,\ldots,Size(A)$\textcolor{gray}{// browse $A$}\\
\indent \indent $A[i].Basin \gets 100 \cdot A[i].Basin/Size(S)$\textcolor{gray}{// translate basins to $\%$ of $S$}\\
\indent \textbf{end for}\\
\indent \textbf{return} $SetNames(A,name,A_{ref})$\textcolor{gray}{// return named attractors, see later}\\
\textbf{end function}\\

\noindent \textbf{function} $ComputeTherapeuticBullets(f_{patho},S,C_{targ},C_{moda},max_{k},\delta,sync,\\ \indent A_{physio},A_{patho},A_{versus})$\\
\indent \textcolor{gray}{// compute a set $B_{therap}$ of therapeutic bullets}\\
\indent $B_{therap} \gets \lbrace \rbrace$\\
\indent $b.Gain[1] \gets Sum(GetCover(A_{physio},A_{patho}))$\textcolor{gray}{// $\bigcup B_{physio,i}$ in $S_{patho}$}\\
\indent \textbf{for} $i_{1} \gets 1,\ldots,Size(C_{targ})$\textcolor{gray}{// browse the target combinations to test}\\
\indent \indent \textbf{for} $i_{2} \gets 1,\ldots,Size(C_{Moda})$\textcolor{gray}{// browse the modality arrangements to test}\\
\indent \indent \indent $b.Targ \gets C_{targ}[i_{1}]$\textcolor{gray}{// the target combination to test}\\
\indent \indent \indent $b.Moda \gets C_{Moda}[i_{2}]$\textcolor{gray}{// the modality arrangement to test}\\
\indent \indent \indent $A_{test} \gets ComputeAttractorSet(f_{patho},S,b,max_{k},1,sync,A_{physio})$\\
\indent \indent \indent $b.Gain[2] \gets Sum(GetCover(A_{physio},A_{test}))$\textcolor{gray}{// $\bigcup B_{physio,i}$ in $S_{test}$}\\
\indent \indent \indent \textbf{if} $IsTherapeutic(b,A_{test},A_{versus},\delta)$\textcolor{gray}{// $b$ is therapeutic}\\
\indent \indent \indent \indent $b.Cover \gets GetCover(A_{physio} \cup A_{versus},A_{test})$\textcolor{gray}{// basins in $S_{test}$}\\
\indent \indent \indent \indent $B_{therap} \gets B_{therap} \cup \lbrace b\rbrace$\textcolor{gray}{// add $b$ to the set of therapeutic bullets}\\
\indent \indent \indent \textbf{end if}\\
\indent \indent \textbf{end for}\\
\indent \textbf{end for}\\
\indent \textbf{return} $B_{therap}$\\
\textbf{end function}\\

\noindent $Sum(container)$ returns the sum of the items in $container$.\\

\noindent \textbf{function} $GetCover(A_{1},A_{2})$\\
\indent \textcolor{gray}{// get the size of the $B_{1,i}$ in $S_{2}$, in $\%$ of $S_{2}$}\\
\indent $cover \gets ()$\\
\indent \textbf{for} $i_{1} \gets 1,\ldots,Size(A_{1})$\textcolor{gray}{// browse the attractors of $A_{1}$}\\
\indent \indent \textbf{if} $\exists i_{2}: A_{2}[i_{2}]=A_{1}[i_{1}]$\textcolor{gray}{// $A_{1}[i_{1}]$ also in $A_{2}$}\\
\indent \indent \indent $cover \gets Append(cover,A_{2}[i_{2}].Basin)$\textcolor{gray}{// get the size of $B_{1,i_{1}}$ in $S_{2}$}\\
\indent \indent \textbf{else}\textcolor{gray}{// $A_{1}[i_{1}]$ not in $A_{2}$}\\
\indent \indent \indent $cover \gets Append(cover,0)$\textcolor{gray}{// then $B_{1,i_{1}}$ is empty in $S_{2}$}\\
\indent \indent \textbf{end if}\\
\indent \textbf{end for}\\
\indent \textbf{return} $cover$\\
\textbf{end function}\\

\noindent $Append(container,item)$ returns $container$ with $item$ added to it.\\

\noindent \textbf{function} $GetVersus(A_{patho})$\\
\indent \textcolor{gray}{// get the pathological attractors}\\
\indent $A_{versus} \gets \lbrace \rbrace$\textcolor{gray}{// the set of the pathological attractors}\\
\indent \textbf{for} $i \gets 1,\ldots,Size(A_{patho})$\textcolor{gray}{// browse the attractors of $A_{patho}$}\\
\indent \indent \textbf{if} $IsSubString(A_{patho}[i].Name,patho)$\textcolor{gray}{// not a physiological attractor}\\
\indent \indent \indent $A_{versus} \gets A_{versus} \cup \lbrace A_{patho}[i]\rbrace$\textcolor{gray}{// then add it to $A_{versus}$}\\
\indent \indent \textbf{end if}\\
\indent \textbf{end for}\\
\indent \textbf{return} $A_{versus}$\\
\textbf{end function}\\

\noindent $IsSubString(s_{1},s_{2})$ returns $true$ if $s_{2}$ is a substring of $s_{1}$.\\

\noindent Remember that $A_{versus}$ is not a true attractor set but the set containing the pathological attractors: $A_{versus} \subset A_{patho}$.\\

\noindent \textbf{function} $GoForward(f,x_{0},b)$\\
\indent \textcolor{gray}{// compute the forward reachable set $fwd$ of $x_{0}$ (asynchronous only)}\\
\indent $fwd \gets \lbrace x_{0}\rbrace$\textcolor{gray}{// $fwd$ contains $x_{0}$ itself}\\
\indent $stack \gets (x_{0})$\textcolor{gray}{// the stack of the states to check, see below}\\
\indent \textbf{for}\\
\indent \indent $x \gets stack[Size(stack)]$\textcolor{gray}{// get the last stack element}\\
\indent \indent $stack \gets stack[1,\ldots,Size(stack)-1]$\textcolor{gray}{// remove the last stack element}\\
\indent \indent $y \gets f(x)$\textcolor{gray}{// prepare all the updated $x_{i}$}\\
\indent \indent \textbf{for} $i \gets 1,\ldots,Size(y)$\textcolor{gray}{// browse the updated $x_{i}$}\\
\indent \indent \indent $z \gets x$\textcolor{gray}{// copy $x$ to preserve its original value}\\
\indent \indent \indent $z[i] \gets y[i]$\textcolor{gray}{// update only one $x_{i}$}\\
\indent \indent \indent $z \gets Shoot(z,b)$\textcolor{gray}{// apply the bullet}\\
\indent \indent \indent \textbf{if} $z \notin fwd$\textcolor{gray}{// new state}\\
\indent \indent \indent \indent $fwd \gets fwd \cup \lbrace z\rbrace$\textcolor{gray}{// then add it to $fwd$}\\
\indent \indent \indent \indent $stack \gets Append(stack,z)$\textcolor{gray}{// and add it to the states to check}\\
\indent \indent \indent \textbf{end if}\\
\indent \indent \textbf{end for}\\
\indent \indent \textbf{if} $Size(stack)=0$\textcolor{gray}{// no new states to visit}\\
\indent \indent \indent \textbf{break}\textcolor{gray}{// so the complete $fwd$ is obtained}\\
\indent \indent \textbf{end if}\\
\indent \textbf{end for}\\
\indent \textbf{return} $fwd$\\
\textbf{end function}\\

\noindent $stack$ is the stack of the visited states for which the possible successors are not yet computed. Once this stack empty, all the visitable states starting from $x_{0}$ are found, that is the complete forward reachable set of $x_{0}$ is obtained.\\

\noindent \textbf{function} $IsTerminal(a,f,b)$\\
\indent \textcolor{gray}{// check if a candidate attractor $a$ is a terminal SCC (asynchronous only)}\\
\indent \textbf{for} $i \gets 1,\ldots,Size(a.States)$\textcolor{gray}{// browse the states of $a$}\\
\indent \indent \textbf{if} $GoForward(f,a.States[i],b) \neq a.States$\textcolor{gray}{// $fwd_{i} \neq a$}\\
\indent \indent \indent \textbf{return} $false$\textcolor{gray}{// then not a terminal SCC}\\
\indent \indent \textbf{end if}\\
\indent \textbf{end for}\\
\indent \textbf{return} $true$\textcolor{gray}{// assumed to be a terminal SCC until proven otherwise}\\
\textbf{end function}\\

\noindent \textbf{function} $IsTherapeutic(b,A_{test},A_{versus},\delta)$\\
\indent \textcolor{gray}{// check if a bullet $b$ is therapeutic}\\
\indent \textbf{if} $b.Gain[2]-b.Gain[1] \geq \delta$\textcolor{gray}{// maybe therapeutic}\\
\indent \indent \textbf{for} $i \gets 1,\ldots,Size(A_{test})$\textcolor{gray}{// browse the attractors of $A_{test}$}\\
\indent \indent \indent \textbf{if} $IsSubString(A_{test}[i].Name,patho) \land A_{test}[i] \notin A_{versus}$\\
\indent \indent \indent \indent \textbf{return} $false$\textcolor{gray}{// because of a \textit{de novo} attractor}\\
\indent \indent \indent \textbf{end if}\\
\indent \indent \textbf{end for}\\
\indent \indent \textbf{return} $true$\textcolor{gray}{// assumed to be therapeutic until proven otherwise}\\
\indent \textbf{else}\\
\indent \indent \textbf{return} $false$\textcolor{gray}{// below the therapeutic threshold}\\
\indent \textbf{end if}\\
\textbf{end function}\\

\noindent \textbf{function} $ReachCycle(f,x_{0},b)$\\
\indent \textcolor{gray}{// compute the cycle reachable from $x_{0}$ (synchronous only)}\\
\indent $cycle \gets (x_{0})$\textcolor{gray}{// begin the trajectory}\\
\indent $x \gets x_{0}$\\
\indent \textbf{for}\\
\indent \indent $x \gets Shoot(f(x),b)$\textcolor{gray}{// update and apply the bullet}\\
\indent \indent \textbf{if} $\exists i: cycle[i]=x$\textcolor{gray}{// cycle found in the trajectory}\\
\indent \indent \indent $cycle \gets cycle[i,\ldots,Size(cycle)]$\textcolor{gray}{// then extract the cycle}\\
\indent \indent \indent \textbf{break}\textcolor{gray}{// mission completed}\\
\indent \indent \textbf{else}\textcolor{gray}{// cycle not yet reached}\\
\indent \indent \indent $cycle \gets Append(cycle,x)$\textcolor{gray}{// then pursue the trajectory}\\
\indent \indent \textbf{end if}\\
\indent \textbf{end for}\\
\indent \textbf{return} $cycle$\\
\textbf{end function}\\

\noindent \textbf{function} $SetNames(A,name,A_{ref})$\\
\indent \textcolor{gray}{// name the attractors of $A$ according to $A_{ref}$, default to $name$}\\
\indent $y \gets A$\textcolor{gray}{// copy $A$ to return a copy}\\
\indent $k \gets 1$\textcolor{gray}{// initiate the default name numbering}\\
\indent \textbf{for} $i \gets 1,\ldots,Size(A)$\textcolor{gray}{// browse the attractors of $A$}\\
\indent \indent \textbf{if} $\exists i_{ref}: A_{ref}[i_{ref}]=A[i]$\textcolor{gray}{// $A[i]$ found in $A_{ref}$}\\
\indent \indent \indent $y[i].Name \gets A_{ref}[i_{ref}].Name$\textcolor{gray}{// then get its name in $A_{ref}$}\\
\indent \indent \textbf{else}\textcolor{gray}{// $A[i]$ not in $A_{ref}$}\\
\indent \indent \indent $y[i].Name \gets CatStrings(name,ToString(k))$\textcolor{gray}{// default name}\\
\indent \indent \indent $k \gets k+1$\textcolor{gray}{// and increment the default name numbering}\\
\indent \indent \textbf{end if}\\
\indent \textbf{end for}\\
\indent \textbf{return} $y$\\
\textbf{end function}\\

\noindent $CatStrings(s_{1},s_{2})$ returns the concatenation of $s_{1}$ and $s_{2}$.\\

\noindent $ToString(item)$ returns the string corresponding to $item$.\\

\noindent This function names the attractors of $A$ according to a reference set $A_{ref}$. If an attractor of $A$ also belongs to $A_{ref}$ then its name in $A_{ref}$ is used, otherwise the default name numbered with $k$ is used.\\

\noindent \textbf{function} $Shoot(x,b)$\\
\indent \textcolor{gray}{// apply a bullet on a state}\\
\indent $y \gets x$\textcolor{gray}{// copy $x$ to return a copy}\\
\indent \textbf{for} $i \gets 1,\ldots,Size(b.Targ)$\textcolor{gray}{// browse the targets}\\
\indent \indent $y[b.Targ[i]] \gets b.Moda[i]$\textcolor{gray}{// apply the corresponding modality}\\
\indent \textbf{end for}\\
\indent \textbf{return} $y$\\
\textbf{end function}\\

\noindent Remember that targets are identified by their position in the state vector, not by their name.\\

\noindent \textbf{function} $Walk(f,x_{0},b,max_{k})$\\
\indent \textcolor{gray}{// perform a random walk of $max_{k}$ steps from $x_{0}$ (asynchronous only)}\\
\indent $x \gets x_{0}$\textcolor{gray}{// start the walk}\\
\indent \textbf{for} $k \gets 1,\ldots,max_{k}$\textcolor{gray}{// for $max_{k}$ steps}\\
\indent \indent $y \gets f(x)$\textcolor{gray}{// prepare all the updated $x_{i}$}\\
\indent \indent $i \gets RandInt(1,Size(x))$\textcolor{gray}{// randomly choose one $x_{i}$}\\
\indent \indent $x[i] \gets y[i]$\textcolor{gray}{// then update the chosen $x_{i}$}\\
\indent \indent $x \gets Shoot(x,b)$\textcolor{gray}{// and apply the bullet}\\
\indent \textbf{end for}\\
\indent \textbf{return} $x$\\
\textbf{end function}\\

\noindent $RandInt(a,b)$ returns a randomly selected integer in $[\![a;b]\!]$ according to a uniform distribution.

\newpage

\section{Appendix 4: case study equations}

\label{equations} Below are the $27$ Boolean equations of the case study derived from the model of bladder tumorigenesis by Elisabeth Remy and colleagues \cite{remy2015modeling}. These equations are also available in text format in the supporting file \texttt{bladder\_equations.txt}. A network-based representation is shown in \hyperref[bladder]{\texttt{Figure \ref*{bladder}}} page \pageref{bladder}.
\begin{small}
\begin{IEEEeqnarray*}{r C l}
AKT&=&PI3K\\
ATM_{lvl1}&=&DNAdamage \land \lnot E2F1_{lvl1} \land \lnot E2F1_{lvl2}\\
ATM_{lvl2}&=&(E2F1_{lvl1} \lor E2F1_{lvl2}) \land DNAdamage\\
CDC25A&=&\lnot CHEK1\slash 2_{lvl1} \land \lnot CHEK1\slash 2_{lvl2} \land \lnot RBL2\\
&&\land (E2F1_{lvl1} \lor E2F1_{lvl2} \lor E2F3_{lvl1} \lor E2F3_{lvl2})\\
CHEK1\slash 2_{lvl1}&=&(ATM_{lvl1} \lor ATM_{lvl2}) \land \lnot E2F1_{lvl1} \land \lnot E2F1_{lvl2}\\
CHEK1\slash 2_{lvl2}&=&(E2F1_{lvl1} \lor E2F1_{lvl2}) \land (ATM_{lvl1} \lor ATM_{lvl2})\\
CyclinA&=&\lnot RBL2 \land \lnot p21CIP \land CDC25A\\
&&\land (E2F1_{lvl1} \lor E2F1_{lvl2} \lor E2F3_{lvl1} \lor E2F3_{lvl2})\\
CyclinD1&=&(RAS \lor AKT) \land \lnot p16INK4a \land \lnot p21CIP\\
CyclinE1&=&\lnot RBL2 \land \lnot p21CIP \land CDC25A\\
&&\land (E2F1_{lvl1} \lor E2F1_{lvl2} \lor E2F3_{lvl1} \lor E2F3_{lvl2})\\
E2F1_{lvl1}&=&\lnot RB1 \land \lnot RBL2 \land ((CHEK1\slash 2_{lvl2} \land ATM_{lvl2} \land \lnot RAS \land E2F3_{lvl1})\\
&&\lor ((\lnot CHEK1\slash 2_{lvl2} \lor \lnot ATM_{lvl2}) \land (RAS \lor E2F3_{lvl1} \lor E2F3_{lvl2})))\\
E2F1_{lvl2}&=&\lnot RBL2 \land \lnot RB1 \land ATM_{lvl2} \land CHEK1\slash 2_{lvl2} \land (RAS \lor E2F3_{lvl2})\\
E2F3_{lvl1}&=&\lnot RB1 \land \lnot CHEK1\slash 2_{lvl2} \land RAS\\
E2F3_{lvl2}&=&\lnot RB1 \land CHEK1\slash 2_{lvl2} \land RAS\\
EGFR&=&(EGFRstimulus \lor SPRY) \land \lnot FGFR3 \land \lnot GRB2\\
FGFR3&=&\lnot EGFR \land FGFR3stimulus \land \lnot GRB2\\
GRB2&=&(FGFR3 \land \lnot GRB2 \land \lnot SPRY) \lor EGFR\\
MDM2&=&(TP53 \lor AKT) \land \lnot p14ARF \land \lnot ATM_{lvl1} \land \lnot ATM_{lvl2} \land \lnot RB1\\
p14ARF&=&E2F1_{lvl1} \lor E2F1_{lvl2}\\
p16INK4a&=&GrowthInhibitors \land \lnot RB1\\
p21CIP&=&\lnot CyclinE1 \land (GrowthInhibitors \lor TP53) \land \lnot AKT\\
PI3K&=&GRB2 \land RAS \land \lnot PTEN\\
PTEN&=&TP53\\
RAS&=&EGFR \lor FGFR3 \lor GRB2\\
RB1&=&\lnot CyclinD1 \land \lnot CyclinE1 \land \lnot p16INK4a \land \lnot CyclinA\\
RBL2&=&\lnot CyclinD1 \land \lnot CyclinE1\\
SPRY&=&RAS\\
TP53&=&\lnot MDM2 \land (E2F1_{lvl2} \lor ((ATM_{lvl1} \lor ATM_{lvl2})\\
&&\land (CHEK1\slash 2_{lvl1} \lor CHEK1\slash 2_{lvl2})))
\end{IEEEeqnarray*}
\end{small}

The four input parameters are $EGFRstimulus$, $FGFR3stimulus$, $GrowthInhibitors$ and $DNAdamage$. The three outputs are evaluated from the returned attractors once the run terminated according to their respective equation:
\begin{small}
\begin{IEEEeqnarray*}{r C l}
Proliferation&=&CyclinE1 \lor CyclinA\\
GrowthArrest&=&p21CIP \lor RB1 \lor RBL2\\
Apoptosis&=&TP53 \lor E2F1_{lvl2}
\end{IEEEeqnarray*}
\end{small}

\newpage

\bibliographystyle{unsrt}
\bibliography{kali_updates}

\end{document}